\documentclass[iop,apj,tighten]{emulateapj}
\usepackage{apjfonts} %If missing fonts, comment out this or google apjfonts to download
\usepackage{amsmath,amstext}
\usepackage{bm}
\usepackage[breaklinks,colorlinks,citecolor=blue,linkcolor=magenta]{hyperref} 
\usepackage[all]{hypcap} %Links go to figures. This breaks deluxetables; use \capstartfalse \capstarttrue around deluxetables to fix it.
\usepackage[usenames,dvipsnames]{color}
\newcommand{\msun}{\ensuremath \mathrm{M}_{\odot}}

\DeclareMathAccent{\dot}    {\mathalpha}{operators}{'137} % The dot is often misplaced otherwise
\DeclareMathAccent{\ddot}    {\mathalpha}{operators}{'177} % same

\newcommand{\be}{\begin{equation}}
\newcommand{\ee}{\end{equation}}
\newcommand{\bdm}{\begin{displaymath}}
\newcommand{\edm}{\end{displaymath}}
\newcommand{\bea}{\begin{eqnarray}}
\newcommand{\eea}{\end{eqnarray}}

\newcommand{\cf}{\textit{cf.}~}

\def\lsim{\lower.5ex\hbox{$\; \buildrel < \over \sim \;$}}
\def\gsim{\lower.5ex\hbox{$\; \buildrel > \over \sim \;$}}

\shorttitle{Testing the binary hypothesis}
\shortauthors{Sesana et al.}

\begin{document}
%test git
\title{Testing the binary hypothesis: pulsar timing constraints on supermassive black hole binary candidates}
\author{Alberto Sesana\altaffilmark{1}, Zolt\'an Haiman\altaffilmark{2,3}, Bence Kocsis\altaffilmark{4} \& Luke Zoltan Kelley\altaffilmark{5}}

\affil{$^1$$^{1}$ School of Physics and Astronomy and Institute of Gravitational Wave Astronomy, University of Birmingham, Edgbaston B15 2TT, United Kingdom}
\affil{$^2$Department of Astronomy, Columbia University, 550 West 120th Street, New York, NY 10027, USA}
\affil{$^3$Department of Physics, New York University, New York, NY 10003, USA}
\affil{$^4$Institute of Physics, E\"otv\"os University, P\'azm\'any P.~s.~1/A, Budapest, 1117, Hungary}
\affil{$^5$Harvard University, Center for Astrophysics, Cambridge, MA 02138, USA}

%#####################################################
\begin{abstract}
  The advent of time domain astronomy is revolutionizing our understanding of the Universe. Programs such as the Catalina Real-time Transient Survey (CRTS) or the Palomar Transient Factory (PTF) surveyed millions of objects for several years, allowing variability studies on large statistical samples. The inspection of $\approx$250k quasars in CRTS resulted in a catalogue of 111 potentially periodic sources, put forward as supermassive black hole binary (SMBHB) candidates. A similar investigation on PTF data yielded 33 candidates from a sample of $\approx$35k quasars. Working under the SMBHB hypothesis, we compute the implied SMBHB merger rate and we use it to construct the expected gravitational wave background (GWB) at nano-Hz frequencies, probed by pulsar timing arrays (PTAs). After correcting for incompleteness and assuming virial mass estimates, we find that the GWB implied by the CRTS sample exceeds the current most stringent PTA upper limits by almost an order of magnitude. After further correcting for the implicit bias in virial mass measurements, the implied GWB drops significantly but is still in tension with the most stringent PTA upper limits. Similar results hold for the PTF sample. Bayesian model selection shows that the null hypothesis (whereby the candidates are false positives) is preferred over the binary hypothesis at about $2.3\sigma$ and $3.6\sigma$ for the CRTS and PTF samples respectively. Although not decisive, our analysis highlights the potential of PTAs as astrophysical probes of individual SMBHB candidates and indicates that the CRTS and PTF samples are likely contaminated by several false positives.
\end{abstract}

%###############################################################################

\maketitle

\section{Introduction}
\label{sec:Introduction}

In the past two decades, it has been realised that supermassive black holes (SMBHs) are a fundamental ingredient of galaxy formation and evolution \citep[see, e.g.,][]{2000MNRAS.311..576K,2006MNRAS.365...11C}, and it is now well established that possibly all massive galaxies host an SMBH at their centre \citep[see][and references therein]{kormendy13}. In the standard $\Lambda$CDM cosmology, structure formation proceeds in a hierarchical fashion, whereby galaxies frequently merge with each other, progressively growing their mass \citep{1978MNRAS.183..341W}. Following the merger of two galaxies, the SMBH hosted in their nuclei sink to the centre of the merger remnant because of dynamical friction (DF), eventually forming a SMBH binary (SMBHB).

The evolution of SMBHBs was first sketched out by \cite{bbr80}. After the initial DF phase, the two SMBHs become bound at parsec scales forming a Keplerian system. At this point DF ceases to be efficient and in the absence of any other physical mechanism at play, the binary would stall. Because the average massive galaxy suffers more than one major merger in its assembly, in this scenario virtually all galaxies would host parsec scale SMBHBs. Galactic nuclei, however, are densely populated with stars and also contain gas. It has been shown that both three-body ejection of ambient stars \citep{2006ApJ...642L..21B,khan11,preto11,2015ApJ...810...49V,2015MNRAS.454L..66S}, interaction with gaseous clumps \citep{2016arXiv160201966G} or with a massive circumbinary disc (\citealt{Escala+2005,MacFadyenMilos2008,hayasaki09,2009MNRAS.393.1423C,2011MNRAS.415.3033R,2011MNRAS.412.1591N,Roedig+2012,Shi+2012,2014ApJ...783..134F,Miranda+2017,Tang+2017}; see also the reviews by \citealt{2012AdAst2012E...3D,Mayer+2013rev}), or interactions between SMBH triplets \citep{Bonetti+2017, Ryu+2017} may provide efficient ways to shrink SMBHBs down to centi-parsec scales, where efficient gravitational wave (GW) emission takes over, leading to final coalescence. Still, binary hardening time-scales can be as long as Gyrs \citep{2015MNRAS.454L..66S,2015ApJ...810...49V}, implying a substantial population of sub-parsec SMBHB lurking in galactic nuclei (see also \citealt{2017MNRAS.464.3131K}).

Sub-pc SMBHBs are extremely elusive objects \citep[see][for recent reviews]{2012AdAst2012E...3D,KomossaZensus2016}. At extragalactic distances, their angular size is well below the resolution of any current instrument, making direct imaging impossible except possibly in radio VLBI observations \citep{DorazioLoeb2017}. Conversely, there is an increasing number of detections of SMBH {\it pairs} (i.e. SMBHs still not gravitationally bound to each others) at hundred pc to kpc separations in merging galaxies, which are their natural progenitors \citep[e.g.][]{2010ApJ...719.1672H,2013ApJ...777...64C}.  With direct imaging impractical, other avenues to observe sub-pc SMBHBs have been pursued, namely spectroscopic identification and time variability. Because of the high ($>1000$ km/s) typical orbital velocities, SMBHBs have tentatively been identified with systems showing significant offsets of the broad-line emission lines compared to the reference provided by narrow emission lines, and/or with frequency shifts of the broad lines over time \citep{2011ApJ...738...20T,2012ApJS..201...23E,Decarli+2013,Ju+2013,Shen+2013,Wang+2017,Runnoe+2017}. The latter, in fact, are generated within the host galaxies at hundred of parsecs from the central SMBHs, whereas the former are generated by gas bound to the SMBHs. If the SMBHs have a significant velocity compared to the galaxy rest frame, and the broad emission region is bound to the individual SMBHs, then broad lines will have an extra redshift/blueshift compared to their narrow counterparts. Note, however, that for very compact SMBHBs those broad lines might in fact be generated within the circumbinary disk \citep{2016MNRAS.459L.124L}, questioning this interpretation. 

With the advent of time domain astronomy, the identification of SMBHBs via periodic variability has been put forward \citep{2009ApJ...700.1952H}. The rationale for selecting candidates in this way is that if gas is being accreted onto a SMBHB, the orbital period of the system could translate into periodic variability of the emitted luminosity. In fact, detailed hydrodynamical simulations show that SMBHBs embedded in circumbinary discs carve a cavity in the gas distribution and the gas streams from the cavity edge onto the binary in a periodic fashion \citep[e.g.][]{al96,MacFadyenMilos2008,hayasaki07,2009MNRAS.393.1423C,2011MNRAS.415.3033R,Shi+2012,Noble+12,Dorazio+2013,2014ApJ...783..134F,ShiKrolik2015}.

Whether such periodic streaming translates into periodicity in the emitted luminosity is less clear. Moreover, it has been pointed out that such periodicity would mostly impact the UV and X-ray emission from the binary, whereas the optical emission coming from the circumbinary disc might be relatively steady \citep{2012MNRAS.420..860S,Farris+2015}, except for the most massive ($M\gsim 10^9\msun$) SMBHBs for which the optical emission arises from gas bound to the individual black holes \citep{2015Natur.525..351D}. Despite these uncertainties, active galactic nuclei (AGN) showing periodic variability are viable candidates for hosting SMBHBs. Based on this hypothesis, \cite{2015MNRAS.453.1562G} (hereafter G15) proposed 111 SMBHB candidates by inspecting the light curves of 243,486 quasars identified in the Catalina Real-time Transient Survey \citep[CRTS,][]{2009ApJ...696..870D}. In a similar effort, \cite{2016MNRAS.463.2145C} (hereinafter C16) identified 33 SMBHB candidates, among 35,383 spectroscopically confirmed quasars in the Palomar Transient Factory \citep[PTF,][]{2009PASP..121.1334R} survey, with somewhat fainter magnitudes and shorter periods than G15. Both groups presented a thorough analysis demonstrating a large excess of periodic sources at odds with the expectations of standard AGN-variability models.

The statistical significance of the detected periodicity depends strongly on the stochastic noise model for underlying quasar variability.  For example, \cite{2016MNRAS.461.3145V} have recently shown that Gaussian
red noise models can naturally lead to frequent false
positives in periodicity searches, especially at inferred periods
comparable to the length of the data stream. At the same time,
observations are expected to yield binaries preferentially at lower
frequencies where they spend the largest fraction of their lifetimes.
On the other hand, pure red noise (i.e a single $f^{-2}$ power-law
noise power spectrum) is a poor description of quasar variability in
general. A better fit to the power spectra observed for large quasar
samples is the damped-random walk (DRW) noise model, in which the red
noise power spectrum flattens to white noise ($\propto f^0$) at low
frequencies (e.g. \citealt{macleod+2010}).  Adopting a DRW reduces the
stochastic noise power at low frequencies, and therefore translate to
a higher significance of an observed periodicity. As a result, the
significance of the inferred periodicities depend strongly on the
poorly constrained underlying noise model (see further discussion of
this point in,
e.g. \citealt{Kelly+2014,2016MNRAS.463.2145C,2016MNRAS.461.3145V,Kozlowski2017}).

In this paper we test the SMBHB hypothesis {\it on physical grounds}. SMBHBs are powerful emitters of nHz gravitational waves (GWs), which are currently probed by pulsar timing arrays \citep[PTAs,][]{1990ApJ...361..300F}. There are three `regional' PTAs currently in operation: the European Pulsar Timing Array \citep[EPTA,][]{2016MNRAS.458.3341D}, the Australian Parkes Pulsar Timing Array \citep[PPTA,][]{2016MNRAS.455.1751R} and the North American Nanohertz Observatory for Gravitational Waves \citep[NANOGrav,][]{2015ApJ...813...65T}. These three collaborations share data under the aegis of the International Pulsar Timing Array \citep[IPTA,][]{2016MNRAS.458.1267V}, with the goal of improving their combined sensitivity. The collective signal from a cosmic population of SMBHBs results in a stochastic GW background \citep[GWB,][]{1995ApJ...446..543R,2003ApJ...583..616J,2003ApJ...590..691W,2008MNRAS.390..192S}, but particularly massive/nearby systems might emit loud signals individually resolvable above the background level \citep{2009MNRAS.394.2255S}.

Recent PTA efforts resulted in several upper limits both for a GWB \citep{2015Sci...349.1522S,2015MNRAS.453.2576L,2016ApJ...821...13A} and for individual sources \citep{2014ApJ...794..141A,2014MNRAS.444.3709Z,2016MNRAS.455.1665B}. Both G15 and C16 showed that each SMBHB candidate identified in CRTS or in PTF is individually compatible with current PTA upper limits. This is partly because timing residuals from those individual SMBHBs are simply too small and partly because their GW emission lies at frequencies above $>10^{-8}$Hz, which is above the PTA detection `sweet spot' at $\approx 10^{-8.3}$Hz. The $10^{-8}$Hz lower limit to the GW frequency, however, is only a selection effect due to the length of the CRTS  and PTF data stream (9 yrs and 4 yrs respectively).

Here we show that, {\it when properly converted into a SMBHB merger rate and extrapolated to lower frequencies, the CRTS and PTF samples are in tension with current PTA measurements.} This is particularly true when virial mass estimates of the candidates are taken at face value; in this case both the CRTS and the PTF sample are severely inconsistent with PTA upper limits. Virial mass estimates are, however, known to be biased high \citep{{2008ApJ...680..169S}}. Correcting for this bias alleviates the inconsistency of the samples with PTA data, but tension persists at $>2\sigma$ level for both. This indicates that both the CRTS and PTF samples are contaminated by several false positives, whose lightcurve variability therefore must have a different physical origin. We show that our conclusions are not severely affected by physical processes potentially capable of suppressing the low frequency GW signal, such as significant eccentricities, or strong environmental coupling.

This paper is organized as follows. We concentrate our investigation on the G15 SMBHB candidate sample, which is presented in \S~\ref{sec2} and used in \S~\ref{sec3} to reconstruct the merger rate of SMBHBs throughout cosmic history. The implied GWB is computed in \S~\ref{sec4} and compared to current PTA upper limits. In \S~\ref{sec5} we apply the same formalism to the C16 sample. The main results are summarized in \S~\ref{sec6}. Throughout this paper, we adopt a concordance $\Lambda$CDM cosmology with $h=0.679,\,\Omega_M=0.306,\,\Omega_\Lambda=0.694$ \citep{2016A&A...594A..13P}.

\newpage
\section{The Catalina survey SMBHB candidate sample}
\label{sec2}
CRTS is a time-domain survey periodically scanning 33,000 deg$^{2}$ (80\% of the whole sky). The data release used by G15 contains the light curves of millions of objects monitored over nine years. Objects have been cross-checked with the 1M quasar catalogue\footnote{http://quasars.org/milliquas.htm} to identify more than 300k spectroscopically confirmed quasars, of which 243,486 have sufficient light curve coverage for a periodicity search. Among these sources, 111 have been flagged as periodically varying and have been proposed as potential SMBHB candidates. For most of these systems, the values of the estimated SMBHB mass, redshift and orbital period is provided by G15 in tabular form.  

\subsection{Observational properties and intrinsic mass estimates}
\label{sec:bias}
%%%%%%%%%%%%%%%%%%%%%%%%%%%%%%%%%%%%%%%%%
\begin{figure}
\includegraphics[scale=0.42,clip=true,angle=0]{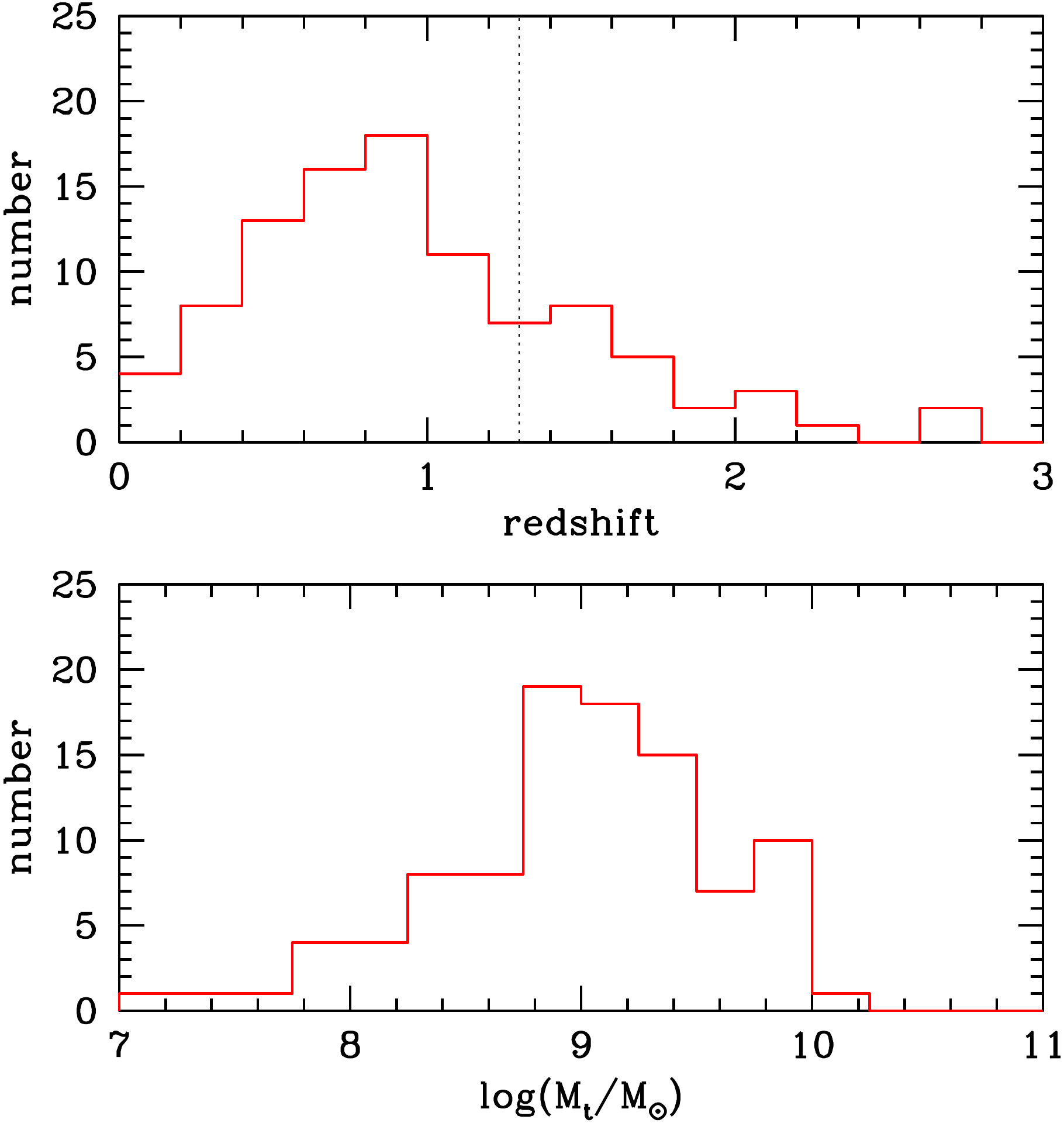}
\caption{Redshift (upper panel) and total mass (lower panel) distribution of the SMBHB candidates identified by \protect\cite{2015MNRAS.453.1562G} in the CRTS. Out of the 111 candidates, only the 98 with reported black hole mass estimates are shown. The dotted vertical line in the upper panel at $z=1.3$ marks the higher redshift considered in SMBHB population models by \protect\cite{2013MNRAS.433L...1S}.}
\label{fig_mz}
\end{figure}
%%%%%%%%%%%%%%%%%%%%%%%%%%%%%%%%%%%%%%%%%
%%%%%%%%%%%%%%%%%%%%%%%%%%%%%%%%%%%%%%%%%
\begin{figure}
\includegraphics[scale=0.42,clip=true,angle=0]{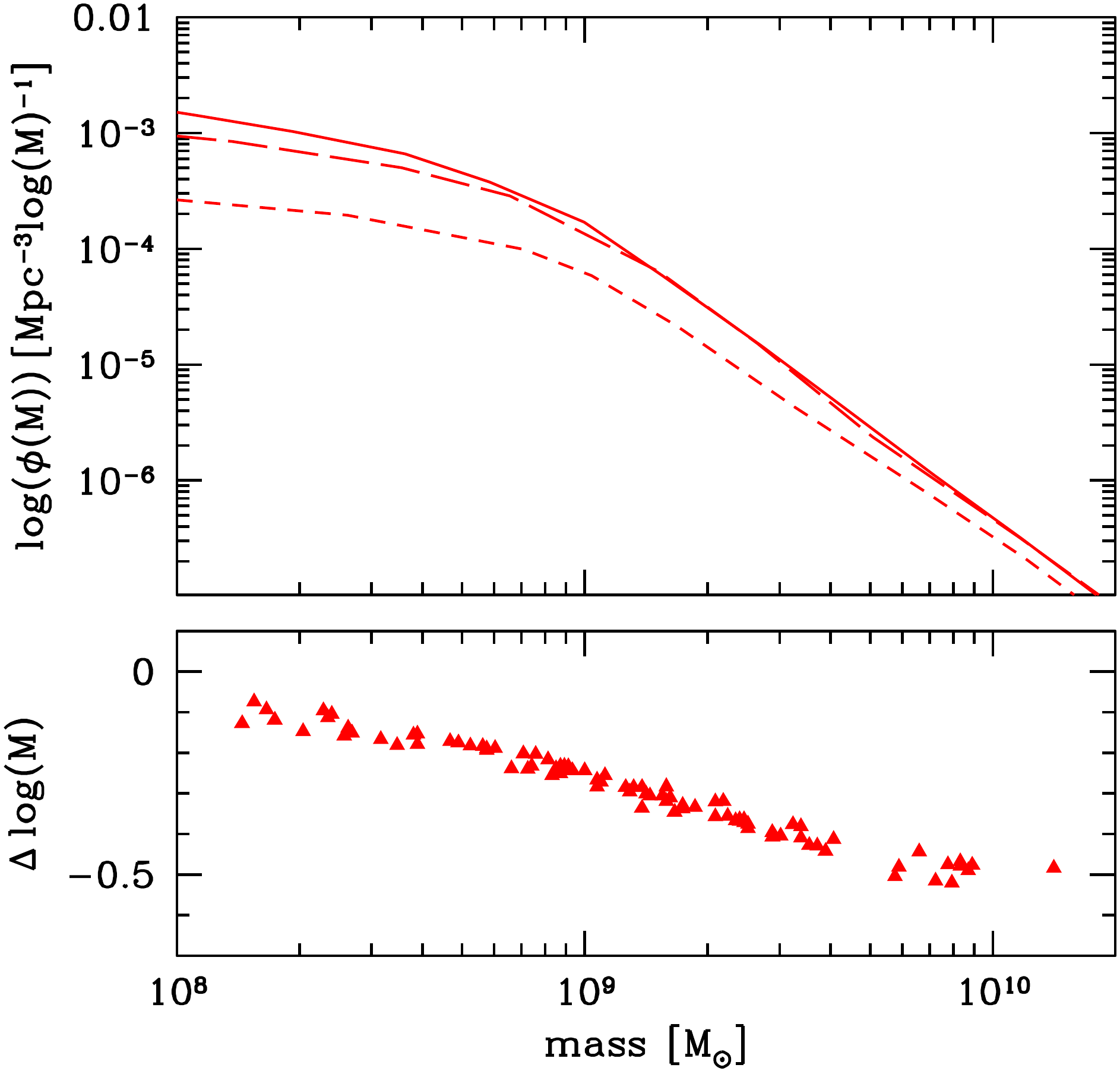}
\caption{Top panel: SMBH mass function derived by  \protect\cite{2007ApJ...654..731H} at $z=0$ (solid), $z=1$ (long--dashed), $z=2$ (short--dashed). Bottom panel: median bias of the virial masses of all the candidates reported in G15 (see main text for full description).}
\label{figbias}
\end{figure}
%%%%%%%%%%%%%%%%%%%%%%%%%%%%%%%%%%%%%%%%%
Of the 111 candidates presented by G15, we consider only the 98 with a reported  measurement of the mass. In the following we will conservatively assume that this is the total SMBHB mass ($M_t=M_1+M_2$). The mass and redshift distributions of these 98 systems, as reported in G15, are plotted in Fig.~\ref{fig_mz}. Interestingly, the number of sources peaks at $z\approx 1$, rapidly declining to zero at $z>2$. In the binary hypothesis, this can be explained by a selection effect. G15's search is sensitive to observed periods between 20--300 weeks and has a fixed magnitude limit. As a result, at higher redshifts, they could only find binaries with shorter rest-frame orbital periods ($<$2 yr at $z>2$), and with increasingly large masses. These massive, compact binaries have decoupled from their circumbinary discs, and are likely well inside the GW-driven inspiral regime  \citep{2009ApJ...700.1952H}, with very short ($\lsim 10^4$yr) inspiral times. As a result, they will be exceedingly rare. Additionally, binaries can take several Gyrs to overcome the `final parsec problem', so that the distribution of SMBHBs might peak at lower redshift with respect to the quasar luminosity function.

In performing our analysis, a correct estimate of the mass of the systems is of paramount importance. The values of $M_t$ reported in G15 and plotted in Fig.~\ref{fig_mz} are taken from the catalog compiled by \cite{2008ApJ...680..169S}, which provides `single epoch mass measurements' using virial BH mass estimators based on the luminosities and the H$\beta$, Mg$_{\rm II}$, and C$_{\rm IV}$ emission lines. We will refer to those mass estimates as {\it virial masses}. Let us consider an object with true mass $M$, with a virial mass estimate $M_e$ and let us define $m={\rm log}M$, and $m_e={\rm log}M_e$. Let us further make the long-standing simplifying assumption \citep[e.g.][]{1913MNRAS..73..359E} that the virial mass estimator, at fixed true mass, follows the log-normal distribution,
%%%%%%%%%%
\begin{equation}
  p(m_e|m)=\frac{1}{\sqrt{2\pi\sigma^2}}\, e^{\frac{(m-m_e)^2}{2\sigma^2}},
  \label{eq:pmeasure}
\end{equation}
%%%%%%%%%%
where $\sigma=0.3$dex is the measured intrinsic scatter. Equation (\ref{eq:pmeasure}) describes the probability of measuring a mass $M_e$ given a true mass $M$. We are, however, interested in an estimate of the true mass $M$, which can be obtained using Bayes theorem
\begin{equation}
  p(m|m_e)\propto p(m_e|m)p(m),
\end{equation}
where $p(m)$ is the prior distribution of the true masses and we omitted a constant normalization factor. In our case $p(m)$ represents the prior knowledge of the SMBH mass function $\phi(m)$, leading to the product
%%%%%%%%%
\begin{equation}
  p(m|m_e)\propto \phi(m) \, e^{\frac{(m-m_e)^2}{2\sigma^2}}.
  \label{eq:mtrue}
\end{equation}
%%%%%%%%%
For a bottom-heavy mass function, this leads to a Malmquist-type bias in the estimate of the true mass $M$, i.e. $\langle m \rangle <  \langle m_e\rangle$ \citep[e.g.][]{1988ApJ...326...19L}. In our calculation we consider the SMBH mass function obtained by \cite{2007ApJ...654..731H}. We note that the normalization of the mass function is irrelevant; our results depend only on the shape of the mass function. The key factor is that $\phi(m)$ is generally described as a broken power-law, with a steep slope $\alpha\approx -2.5$ at the high mass end. In that mass range, when combined with $\phi(m)$, a measured $m_e$ implies an underlying true mass $m$ which is approximately 0.5dex smaller \citep[see][]{2008ApJ...680..169S}. This is of capital importance, because it means that at the high mass end, the $M_t$ values reported in G15 are biased high by a factor of $\approx 3$. This is illustrated in Fig.~\ref{figbias}, where the mass function at different redshifts is plotted along with the median bias $\Delta{\rm log}(M)$ computed by taking the difference between the virial masses reported in G15 and the median of the true mass distribution obtained via equation (\ref{eq:mtrue}). We checked that equation (\ref{eq:mtrue}) implies a bias of 0.55dex when combined with a single power-law with $\alpha\approx -2.6$, consistent with the calculation in \cite{2008ApJ...680..169S}. We also checked that our results are robust against the specific $\phi(m)$ choice: using mass functions derived by \cite{2012ApJ...746..169S} and \cite{2009ApJ...692.1388K} yields quantitatively similar results. This is not surprising, since all mass functions share a consistent steep decay at the high mass end. In the following, we therefore consider two models:
\begin{itemize}
\item {\bf Model~True}: `true' masses are generated by drawing randomly from the probability distribution in equation (\ref{eq:mtrue}) where $\phi(m)$ is taken from \cite{2007ApJ...654..731H}. This is our fiducial model.
\item {\bf Model~Vir}: we assume virial masses given by \cite{2008ApJ...680..169S} and reported in G15 are unbiased and we apply a further uncertainty of $\sigma=0.3$dex to the measured value.
\end{itemize}

The second scenario is included for two reasons. First, comparing the
`True' and the `Vir' models allows us to quantify the importance
of correcting for a systematic bias in the virial mass estimates. This
correction depends on the assumed shape of the intrinsic mass
function, and is often neglected in the literature.  Second, the
approach in the `True' models further assumes that the scatter
observed in the virial mass estimates, measured in practice at fixed
luminosity ($L_\lambda$) and line width (FWHM of broad lines),
represents the scatter at a fixed true mass.  This is the most common
interpretation, which leads to equation~(\ref{eq:mtrue}) (see,
e.g.~\citealt{2008ApJ...680..169S}).  However, we note that virial
mass estimates $M_e$ are calibrated against masses $M_r$ determined
from reverberation mapping for a subset of quasars
(e.g.~\citealt{Peterson2014}).
If one makes the extreme assumption that the reverberation masses are
indeed the true masses, with no errors, then the scatter observed in
the 2D plane of ($M_e,M_r$), measured with the respect to the line
$M_e=M_r$, can be interpreted as the scatter in the true mass $M_r$ at
fixed $M_e$.  In this extreme limiting case, the `Vir' models would
give the correct probability distribution of the true masses.

%%%%%%%%%%%%%%%%%%%%%%%%%%%%%%%%%%%%%%%%%%%%%
\subsection{Assigning individual SMBH masses}

The procedure detailed above provides only an estimate of the true total mass of each individual SMBHB candidate. To compute the associated GW signal the mass ratio $q=M_2/M_1$ (where by definition $M_1>M_2$) of the sources is also needed. Inferring the $q$ distribution of observable SMBHBs is far from being a trivial task, and we will see that it is an important factor in assessing PTA constrains. We describe in the following some of the subtleties that come into play, which led us to consider three different scenarios.

%%%%%%%%%%%%%%%%%%%%%%%%%%%%%%%%%%%%%%%%%
\begin{figure}
\includegraphics[scale=0.42,clip=true,angle=0]{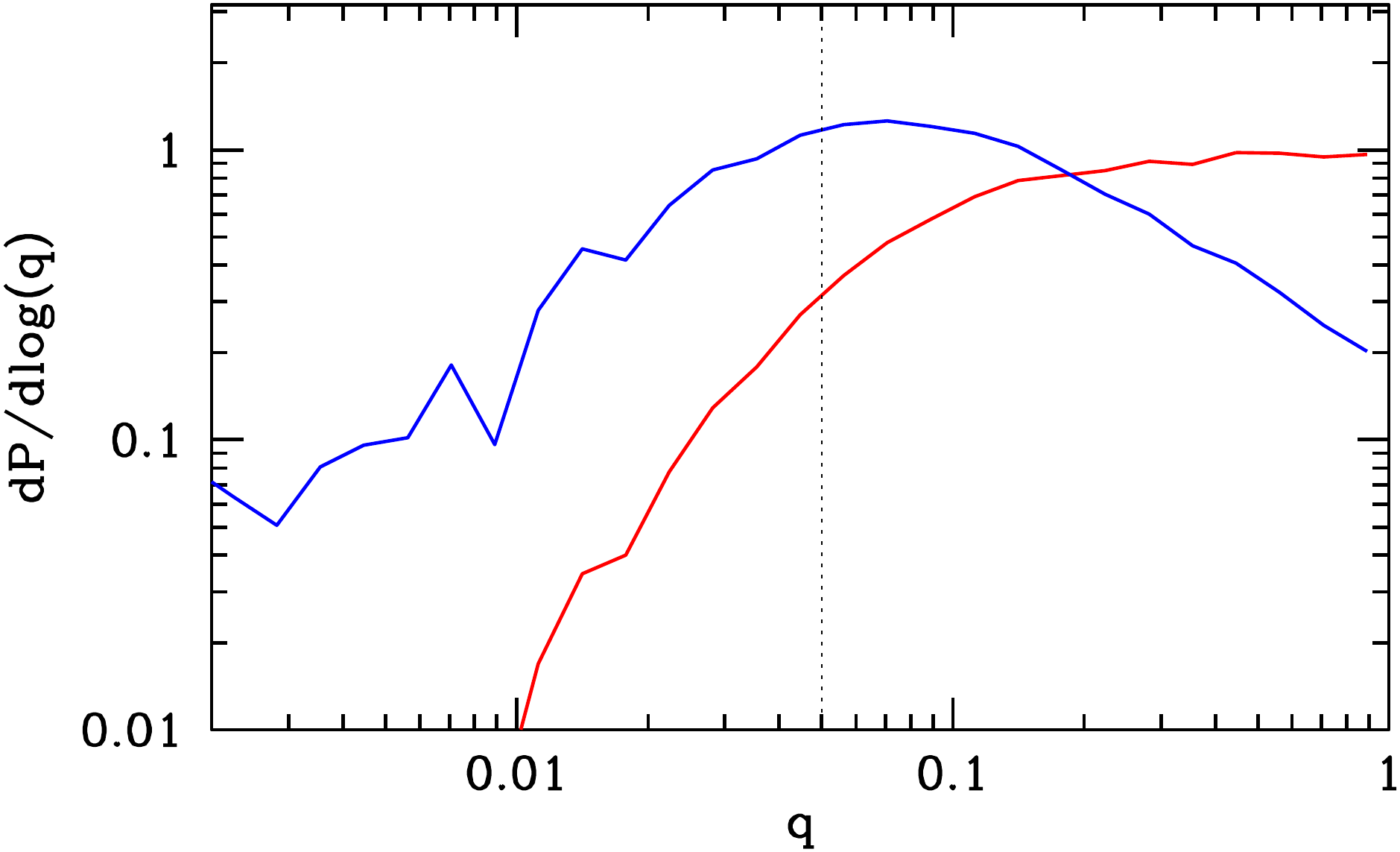}
\caption{Probability distribution of the mass ratio of merging SMBHBs in the Millennium Run considering only systems with $M_t>3\times 10^8\msun$ merging at $z<2$ (red line) compared to the $q$ probability distribution of a SMBHB observed in a selected frequency range (blue line). The vertical dotted line marks $q=0.05$, above which accretion-induced periodicity is seen in numerical simulations of SMBHBs in circumbinary disks (see text for details).}
\label{figqdist}
\end{figure}
%%%%%%%%%%%%%%%%%%%%%%%%%%%%%%%%%%%%%%%%%

In comparing their sample with PTA upper limits, G15 assumed a typical $q=1/2$. Although this might appear a reasonable choice, galaxy formation models predict indeed a larger range of $q$, as shown in Fig.~\ref{figqdist}. Here we consider all the $N_m$ SMBHBs with $M_t>3\times 10^8\msun$ merging at $z<2$ in a modified version of the galaxy formation models of \cite{2013MNRAS.428.1351G} implemented in the Millennium Simulation \citep{2005Natur.435..629S}. Modifications included populating merging galaxies with SMBHBs according to a specific scaling relation \citep[in this case the M--$\sigma$ relation from][]{2009ApJ...698..198G}, and adding appropriate delays between galaxy mergers and SMBHB mergers to account for dynamical friction and SMBHB hardening. Plotted is the probability distribution ${\cal P}({\rm log}q)$ (normalized so that $\int {\cal P}({\rm log}q)d{\rm log}q=1$) of those merging systems. The distribution is essentially flat down to $q=0.1$, and drops at lower mass ratios. We see, however, that mergers down to $q\approx0.01$ are still possible. Similar distributions, albeit sometimes flatter and extending to even lower $q$, have been found in semi-analytic models \citep[e.g.][]{2012MNRAS.423.2533B} and are produced by full cosmological hydrodynamical simulations like Illustris \citep{2017MNRAS.464.3131K}. 

If we take a snapshot of the sky, the $q$ distribution of SMBHB observed {\it at a given frequency} is not the same as that of the merging systems. In fact, if we assume for simplicity that all binaries are circular and purely GW driven, the residence time at a given frequency is proportional to \citep[see, e.g][]{2008MNRAS.390..192S},
\begin{equation}
  \label{dtdf}
  \frac{dt}{df}=\frac{5}{96\,\pi^{8/3}}\left(\frac{G{\cal M}}{c^3}\right)^{-5/3}f^{-11/3},
\end{equation}
where ${\cal M}=M_1^{3/5}M_2^{3/5}/M_t^{1/5}=M_1q^{3/5}/(1+q)^{1/5}$ is the binary chirp mass. Therefore, for a given $M_1$ (and ignoring the $1+q\approx 1$ factor) $dt/df\propto q^{-1}$. Since the number of observable binaries in a given frequency window is $dN/df\propto N_m\times dt/df$, their actual $q$ distribution is proportional to ${\cal P}({\rm log}q)/q$. This is shown by the blue line in Fig.~\ref{figqdist}: the $q$ distribution is now skewed towards lower values, peaking at $q\lesssim0.1$. Longer-lived, low-$q$ binaries should therefore be quite common. Whether they are observable as periodic quasars, however, is less clear. In the standard picture, periodicity in the light curve is associated with variability in the accretion onto a SMBHB as $L(t)\propto \dot M(t)$. The typical variability of $0.2$--$0.5$ magnitudes observed in CRTS thus corresponds to accretion fluctuations (and relative luminosity fluxes) of about $20\%$--$60\%$, necessitating a relatively high $q$. For example, \cite{2014ApJ...783..134F} and \cite{Dorazio+2016} find that $q>0.05$ is needed to get a distinctive variability pattern in the accretion rate (which is marked in Fig.~\ref{figqdist}). On the other hand, \cite{2015Natur.525..351D} proposed that the sinusoidal behaviour of the light curve is given by relativistic Doppler boosting.
This explanations is preferred at low mass ratios ($q\lsim 0.1$),
because low $q$ (i) increases the secondary's orbital velocity and the
amplitude of the Doppler modulation, while (ii) reducing the
hydrodynamical variability. Thus, the Doppler variability is
complementary to the accretion-induced variability.
Besides the technical details of the source of variability, the large accretion rates implied by the CRTS sample of bright quasars are generally found to be triggered by major mergers \citep[e.g.][]{2006MNRAS.365...11C}. Therefore the $q$ distribution of SMBHBs hosted in bright quasars might be biased high.

In light of these uncertainties, we construct three mass ratio models:
\begin{itemize}
\item {\bf Model~hiq}: $q$ drawn from a log-normal distribution with $\sigma_{{\rm log}q}=0.5$dex, peaked at ${\rm log}q=0$ (and we consider only ${\rm log}q<0$). This model is representative of a biased high $q$ distribution, and is similar to the $q=1/2$ study case considered in G15.
\item {\bf Model~fid}: $q$ is drawn from the distribution plotted in Fig.~\ref{figqdist} (blue curve), but with a minimum cut-off placed at $q=0.05$. This model is inspired by accretion-induced periodicity favouring higher $q$.
\item {\bf Model~loq}: $q$ is drawn from the distribution plotted in Fig.~\ref{figqdist} (blue curve), without any cut-off. This model preserves binaries with $q\lesssim 0.1$ and is inspired by the Doppler boosting scenario, favouring lower $q$.
\end{itemize}
Coupled with the two mass-models described in the previous section, this gives us a total of six models that we label {\bf Model\_True\_hiq, Model\_True\_fid, Model\_True\_loq, Model\_Vir\_hiq, Model\_Vir\_fid, Model\_Vir\_loq}. `True' models, that correct for the intrinsic bias in virial mass measurement, are our default models and the following discussion will concentrate on them. Among those we pick Model\_True\_fid as our fiducial model, since it combines corrected mass estimates with a mass ratio distribution derived from cosmological simulations coupled with a minimum cut which is motivated by systematic hydrodynamical simulations. We stress once again, however, that our fiducial $q$ distribution has been
derived by considering {\it all} SMBHB mergers found in the
Millennium Simulation. The CRTS sample, however, is composed by
bright quasars, which are generally associated with major mergers. It
is therefore quite possible that the expected $q$ distribution of the
CRTS sample is skewed towards higher $q$ with respect to our fiducial
model, making our assessments conservative.

Note that Model\_True\_loq and Model\_Vir\_loq imply a moderate inclination with respect to the observer line of sight \citep[otherwise the boosting would not be efficient enough, see details in][]{2015Natur.525..351D}. This implies that the CRTS observed sources would be an incomplete sample (including only systems with favourable inclination) of the underlying SMBHB population. We do not attempt to address this incompleteness in the following analysis, which makes our estimates for those two models conservative, and accounts for the possibility of accretion-induced variability at low $q$ \citep{2014ApJ...783..134F}.

%%%%%%%%%%%%%%%%%%%%%%%%%%%%%%%%%%%%%%%%%
\section{Building mock SMBHB populations}
\label{sec3}

For each of the six models enumerated in the previous section, we take the 98 CRTS candidates with estimated total mass and we assign them $M_t$ and $q$ by drawing from the respective distributions. We repeat the procedure 1000 times, to get a statistical ensemble of SMBHBs samples under each of the model prescriptions. 

\begin{figure}
\includegraphics[scale=0.44,clip=true,angle=0]{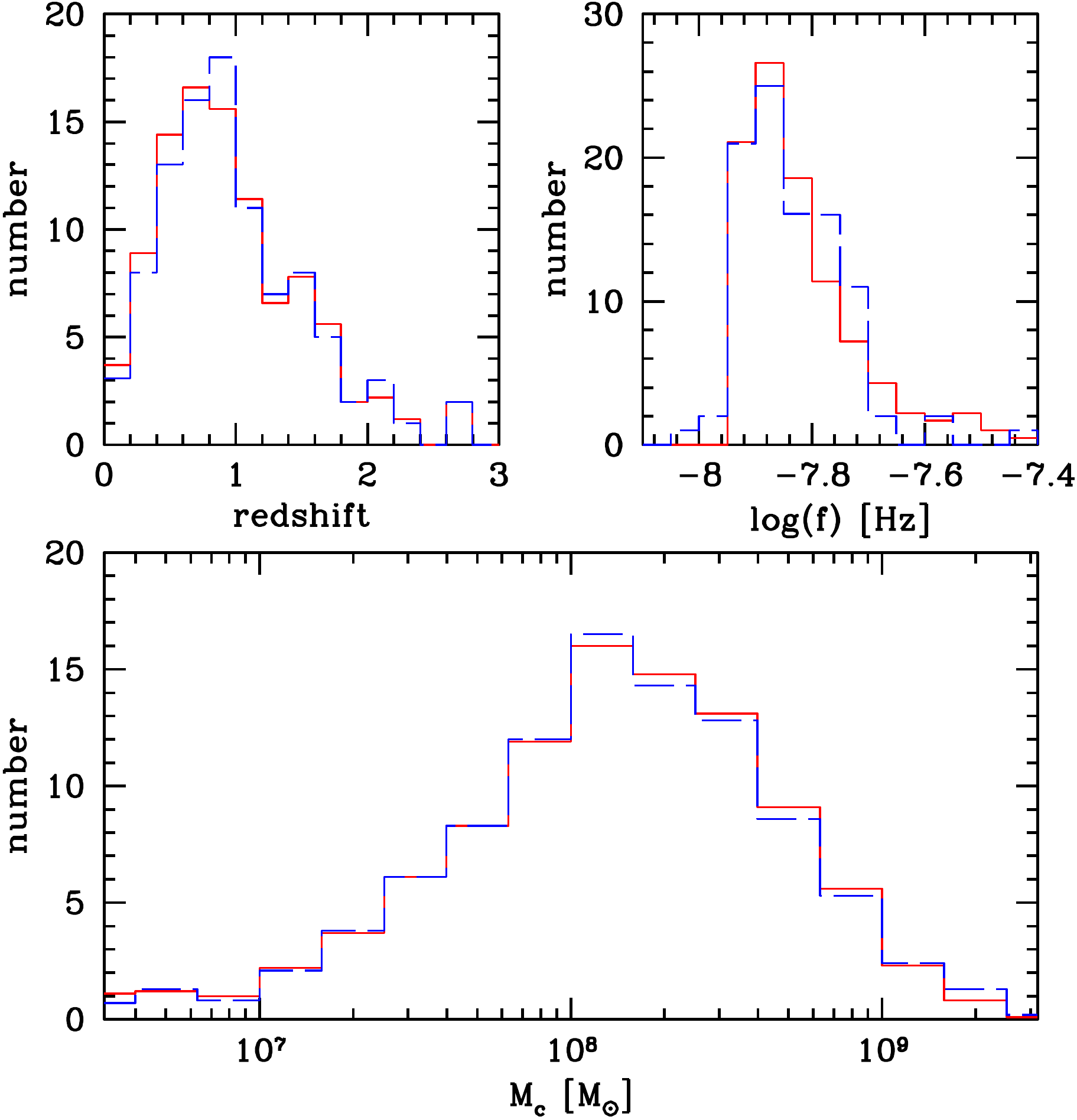}
\caption{Comparison between the 98 SMBHB candidates in the CRTS, and a mock population of SMBHBs constructed from the merger rate implied by the CRTS systems. {\em Upper left:} redshift distribution. {\em Upper right:} rest-frame frequency distribution. {\em Lower panel:} chirp mass distribution. The chirp mass distribution is constructed assuming Model\_True\_fid (see text for details). In all panels, solid red histograms are the CRTS sample, dashed blue histograms are an average over 10 Monte Carlo realisations of Model\_True\_fid.} 
\label{fig_comp}
\end{figure}
%%%%%%%%%%%%%%%%%%%%%%%%%%%%%%%%%%%%%%%%%
Each individual realisation of the 98 CRTS SMBHB candidates can be used to infer a SMBHB merger rate as follows. Neglecting, for the moment, completeness issues, the output of the CRTS can be treated as an all sky population of binaries with the given masses and redshifts, emitting in some frequency range. We note that, because of the limited time span of the data, CRTS is sensitive only to SMBHBs with {\it observed} orbital periods shorter than the threshold value $\tilde{P}_{\rm orb}=6$yrs. This corresponds to a {\it rest frame} GW frequency $\tilde{f}_r=2\times \tilde{P}_{\rm orb}^{-1}(1+z)$. The rest frame coalescence time-scale of a (circular, GW driven) SMBHB emitting at $\tilde{f}_r$ is \citep{1964PhRv..136.1224P}  
\begin{equation}
  T_c(\tilde{f}_r)=\frac{5}{256}\left(\frac{G{\cal M}}{c^3}\right)^{-5/3}(\pi\tilde{f}_r)^{-8/3}.
  \label{eq:tc}
\end{equation}
Suppose now we have $N$ identical (in mass and redshift) binaries emitting at frequencies $f_r>\tilde{f}_r$ identified in the CRTS sample. By virtue of the continuity equation for SMBHB evolution, we can convert this number, into a merger rate using:
\begin{equation}
\label{dndt}
\dot{N}\equiv\frac{dN}{dt_r}=\frac{N(f_r>\tilde{f}_r)}{T_c(\tilde{f}_r)}.
\end{equation}
We stress once again that $T_c(\tilde{f}_r)$ is the coalescence time at the longest orbital period probed by CRTS ($\tilde{P}_{\rm orb}$), which is the time-scale defining the sample population, and not at the period of each specific binary candidate. We can then generalize the argument to a distribution of SMBHBs with different masses and redshifts and numerically construct a binned distribution $\Delta^2\dot{N}/(\Delta{\cal M}\Delta{z})$ by summing up the sources in the CRTS sample as follows:
\begin{equation}
\frac{\Delta^2{\dot{N}}}{\Delta{\cal M}\Delta{z}}=\sum_{i\in\Delta{\cal M}}\sum_{i\in\Delta{z}}\frac{1}{{\Delta{\cal M}\Delta{z}}T_{c,i}(\tilde{f}_r)}.
  \label{eq:dndtr}
\end{equation}
Here $i=1,...,98$ is an index identifying the SMBHB candidates, and $T_{c,i}(\tilde{f}_r)$ is the coalescence time-scale of the $i-$th binary according to equation (\ref{eq:tc}). Note that although equation (\ref{eq:dndtr}) depends on the choice of the bins $\Delta{\cal M}$ and $\Delta{z}$, when computing the total merger rate via equation (\ref{eq:dndt}) or the GW signal via equation (\ref{hcdE}) below, we integrate over $d{\cal M}$ and $dz$, and the final results are bin-independent. To ease the notation, we switch to the continuum and use the differential form $\Delta^2\dot{N}/(\Delta{\cal M}\Delta{z})\rightarrow d^3N/(d{\cal M}dz dt_r)$. We stress, however that all computations have been performed numerically on binned distributions and the robustness of the results have been checked against the choice of bin sizes.

To check whether our merger rate is consistent with the observed CRTS SMBHB population, we can now construct the expected $d^3N/(d{\cal M}dzdf_r)$ as
\begin{equation}
  \frac{d^3N}{d{\cal M}dz df_r}=\frac{d^3N}{d{\cal M}dz dt_r}\frac{dt_r}{df_r},
  \label{eq:dndfr}
\end{equation}
where $dt_r/df_r$ is the same as equation (\ref{dtdf}) but evaluated in the source rest-frame. We can then perform a Monte Carlo sampling of the distribution given by equation (\ref{eq:dndfr}) and compare it to the original CRTS SMBHB candidate ensemble. The comparison is given in Fig.~\ref{fig_comp}, where we show the chirp mass, rest-frame frequency and redshift distribution averaged over 10 realisations of Model\_True\_fid. The inferred merger rate numerically constructed as explained above is perfectly consistent with the observed CRTS SMBHB population. Equation (\ref{eq:dndtr}) therefore provides a reliable estimate of the SMBHB merger rate implied by the observed candidates. We caution that this might differ from the {\it intrinsic} SMBHB merger rate as, in practice, completeness may depend on $M$, $q$, $f_r$, which distorts the distribution as discussed below.

\subsection{Coalescence rates}
\label{coalescence_rate}
%%%%%%%%%%%%%%%%%%%%%%%%%%%%%%%%%%%%%%%%%
\begin{figure}
\includegraphics[scale=0.42,clip=true,angle=0]{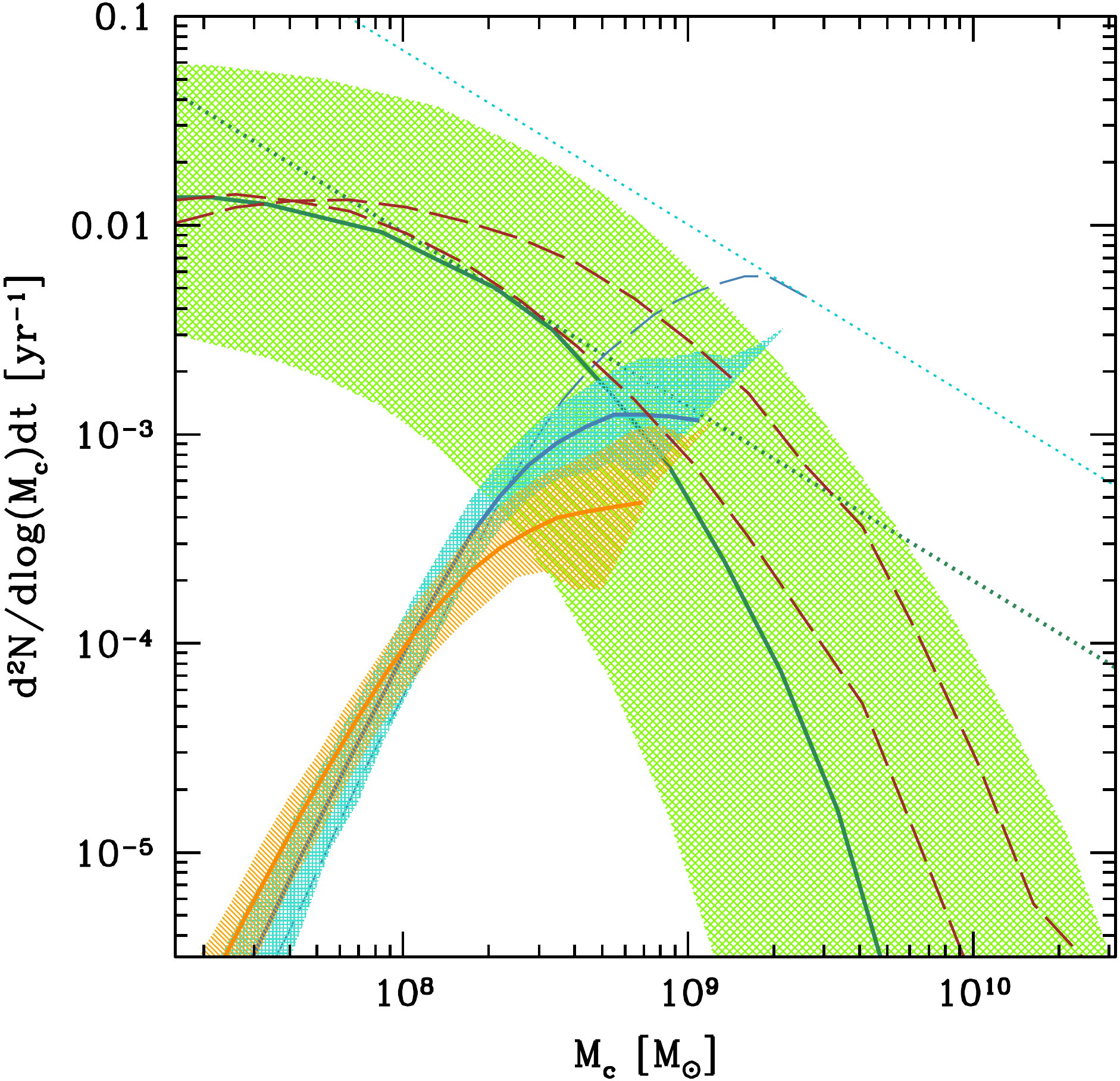}
\caption{Merger rate mass function comparison. The green shaded area is the 95\% interval produced by theoretical models presented in \protect\cite{2013MNRAS.433L...1S} and the solid dark green line is the median value. The blue and orange shaded  areas are the 95\% intervals based on the observed samples of periodic AGNs produced by Model\_True\_hiq and Model\_True\_loq described in the text, and the thick blue and orange lines are the respective median values. Model\_True\_fid lies in between the two and it is omitted to preserve figure readability. We also show the median merger rate obtained in Model\_Vir\_hiq (blue long-dashed line). The two dashed brown lines are two models selected from \protect\cite{2009MNRAS.394.2255S}. The dotted curves with a $-5/6$ slope are tangent to the median values predicted by selected models and show which chirp mass contributes the most to the GW signal (see text for details).}
\label{figmf}
\end{figure}
%%%%%%%%%%%%%%%%%%%%%%%%%%%%%%%%%%%%%%%%%
%%%%%%%%%%%%%%%%%%%%%%%%%%%%%%%%%%%%%%%%%
\begin{figure}
\includegraphics[scale=0.44,clip=true]{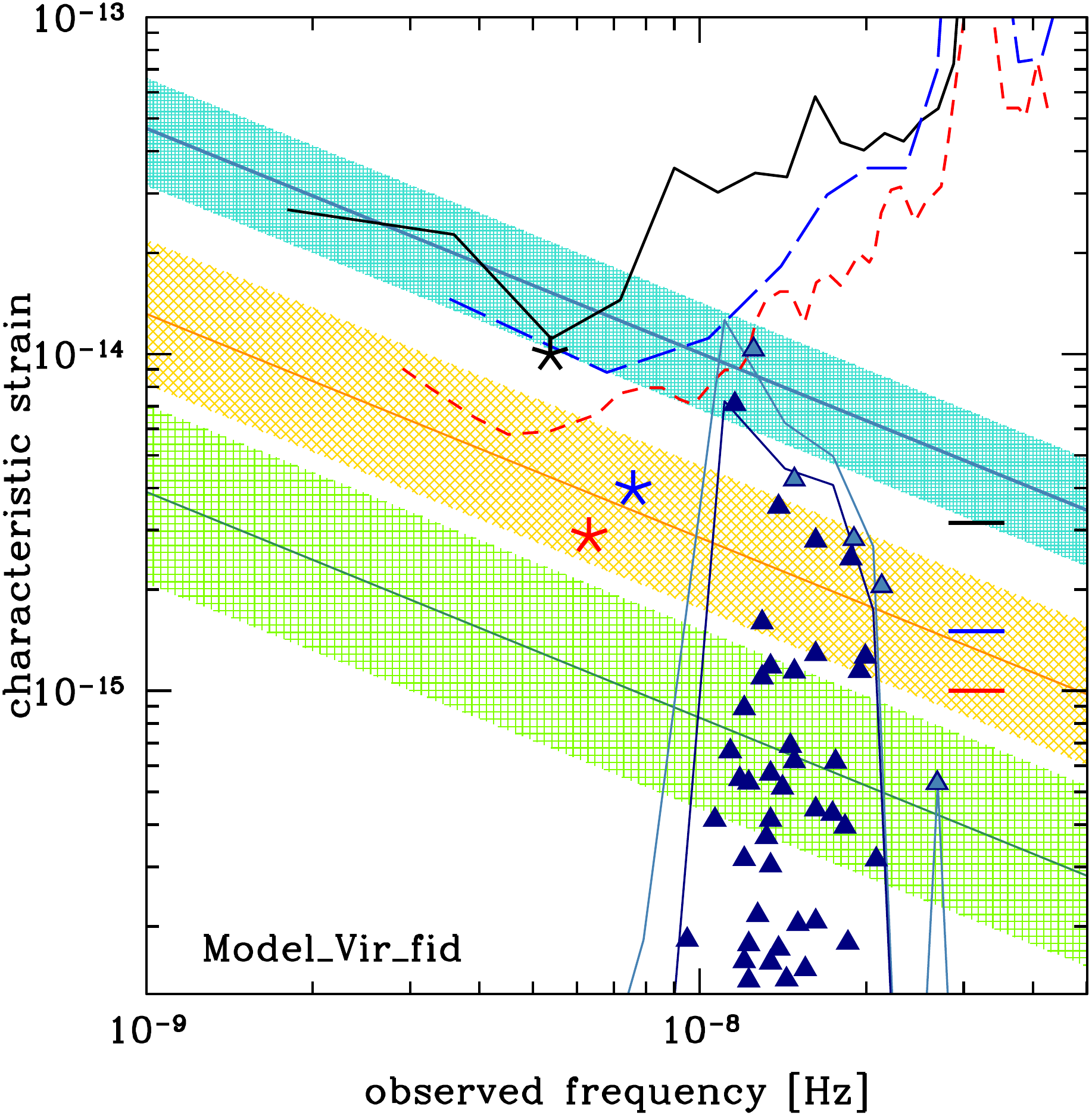}
\caption{Example of the characteristic amplitude of the GW signal for circular, GW-driven binaries. The orange and green shaded areas are 68\% intervals produced by selected models presented in \protect\cite{2013MNRAS.433L...1S} and \protect\citep{2016MNRAS.463L...6S} (see text for details); the solid orange and green lines are their respective median values. The blue shaded area is the 68\% interval produced by Model\_Vir\_fid and the solid blue line is the median value. The dark blue triangles illustrate one Monte Carlo realization of individual GW signals from the G15 sample.  The light blue triangles are sources that are individually louder than all other binaries in the same frequency bin  (and therefore, potentially resolvable individually by a putative PTA with enough sensitivity), taken here to be $\Delta f=3.17$nHz (i.e.~10yr$^{-1}$). The corresponding light blue line is the resulting overall signal, and the dark blue line is the level of the GWB after these putative resolvable sources are subtracted. The solid black, long-dashed blue and short-dashed red lines represent the sensitivities to a stochastic GW signal as a function of frequency for the EPTA, NANOGrav and PPTA respectively. The corresponding stars represent the 95\% upper limit to the integrated GWB from each PTA, placed at the frequency at which it is most sensitive. The horizontal ticks are the extrapolation of those limits to a frequency of 1yr$^{-1}$.}
\label{fighc}
\end{figure}
%%%%%%%%%%%%%%%%%%%%%%%%%%%%%%%%%%%%%%%%%

We can now compute the {\it observed} differential merger rate per unit chirp mass, by integrating equation (\ref{eq:dndtr}) in a given redshift range,
%%%%%%%%%%%%
\begin{equation}
\frac{d^2N}{d{\cal M}dt}=\int_{z_{\rm min}}^{z_{\rm max}}\frac{d^3N}{d{\cal M}dz dt_r}\frac{dt_r}{dt}dz.
  \label{eq:dndt}
\end{equation}
%%%%%%%%%%%%
Results are shown in Fig.~\ref{figmf}, where the merger rates of Model\_True\_hiq and Model\_True\_loq are integrated in the redshift range $0<z<1.3$ and are compared to theoretical estimates from the literature (integrated in the same redshift range). In particular, we consider an updated version of the observation-based models of \cite{2013MNRAS.433L...1S}, and two models extracted from the Millennium simulation \citep{2005Natur.435..629S} and described in \cite{2009MNRAS.394.2255S}. The sharp decline at ${\cal M}<3\times10^8\msun$ of Model\_True\_hiq and Model\_True\_loq is likely due to CRTS incompleteness; lighter SMBHBs are intrinsically fainter and might be missing in the relatively shallow Catalina survey. Both models are consistent with theoretical estimates, even though incompleteness of the CRTS sample can significantly increase the underlying merger rate which might create tension at high masses, especially for Model\_True\_hiq. We will discuss the effect of incompleteness on our results in \S~\ref{sec:incompleteness}.

Fig.~\ref{figmf} also shows the median merger rate obtained when Model\_Vir\_hiq is assumed (blue long-dashed line). In this case the CRTS population is inconsistent at least at a 95\% level at ${\cal M}>2\times10^9\msun$, with respect to expectations from hierarchical clustering. Note that incompleteness of the CRTS sample can only increase the merger rate which makes this discrepancy worse at high masses; in fact, although undetected low mass systems might close the gap at ${\cal M}<10^8\msun$, any correction for incompleteness can only make the mismatch at the high mass end more severe. In particular, we will see in equation (\ref{hcdE}) below that for a given merger rate, the gravitational wave strain is $h_c\propto {\cal M}^{5/6}$. This means that the largest contribution to the GW signal comes from the value of ${\cal M}$ where the merger rate distribution is tangent to a line with $-5/6$ slope in the log-log plane, as depicted in Fig.~\ref{figmf}. We see that the line tangent to the median of Model\_Vir\_hiq gives a systematically higher rate by an order of magnitude than the merger rate estimates from theoretical models, which results in a large over prediction of the GW signal (as later shown in Fig.~\ref{fig_a}). This highlights that a proper estimate of the true masses of the CRTS candidates is of paramount importance.  

\section{PTA implications of the Catalina sample}
\label{sec4}

With a reliable estimate of the SMBHB merger rate in hand, we can now proceed to the computation of the expected GW signal. We start by considering circular, GW-driven binaries. Following \cite{2001astro.ph..8028P} and \cite{2008MNRAS.390..192S}, we can write the overall GW signal as:
%%%%%%%%%%%%%%%%%%%%%%%%%%%%%%%%%
\begin{equation}
h_c^2(f) =\frac{4}{\pi f^2}\int_0^{\infty} 
{dz\int_0^{\infty}d{\cal M} \, \frac{d^2n}{dzd{\cal M}}
{1\over{1+z}}~{{dE_{\rm gw}({\cal M})} \over {d\ln{f_r}}}}\,.
\label{hcdE}
\end{equation}
%%%%%%%%%%%%%%%%%%%%%%%%%%%%%%%%%
The energy radiated per logarithmic frequency interval is
%%%%%%%%%%%%%%%%%%%%%%%%%%%%%%%%%
\begin{equation}
\frac{dE_{\rm gw}}{d\ln{f_r}}=\frac{\pi^{2/3}}{3}{\cal M}^{5/3}f_r^{2/3},
\label{dedlnf}
\end{equation}
%%%%%%%%%%%%%%%%%%%%%%%%%%%%%%%%%
and ${d^2n}/{dzd{\cal M}}$ is the comoving number density of mergers per unit redshift and chirp mass and is related to the overall cosmic merger rate of equation (\ref{eq:dndtr}) via
%%%%%%%%%%%%%%%%%%%%%%%%%%%%%%%%%
\begin{equation}
  \frac{d^2n}{dzd{\cal M}}=\frac{d^3N}{d{\cal M}dz dt_r}\frac{dt_r}{dz}\frac{dz}{dV_c}.
  \label{eq:d2ndzdm}
\end{equation}
%%%%%%%%%%%%%%%%%%%%%%%%%%%%%%%%%
The functions $dt_r/dz$ and $dz/dV_c$ are given by standard cosmology. We can therefore construct the expected GW signal by using equation (\ref{hcdE}) for all our models. For each model, we compute $h_c$ from the 1000 realisations of the CRTS ensemble, to get a measurement of the uncertainty in the predicted signal amplitude.

Note that in the  circular GW-driven scenario, the GW signal computed through equation (\ref{hcdE}) is a simple power-law with spectral slope $-2/3$ \citep{2001astro.ph..8028P}. The GW amplitude is thus customarily written as
%%%%%%%%%%%%%%%%%%%%%%%%%%%%%%%%%
\begin{equation}
  h_c=A\left(\frac{f}{1{\rm yr}^{-1}}\right)^{-2/3},
  \label{hc:par}
\end{equation}
%%%%%%%%%%%%%%%%%%%%%%%%%%%%%%%%%
and PTA results are quoted as limits on the amplitude normalization $A$ at a nominal frequency of 1yr$^{-1}$.

\subsection{Model\_Vir: severe inconsistency with PTA measurements}
%%%%%%%%%%%%%%%%%%%%%%%%%%%%%%%%%%%%%%%%%
\begin{figure}
\includegraphics[scale=0.53,clip=true,angle=0]{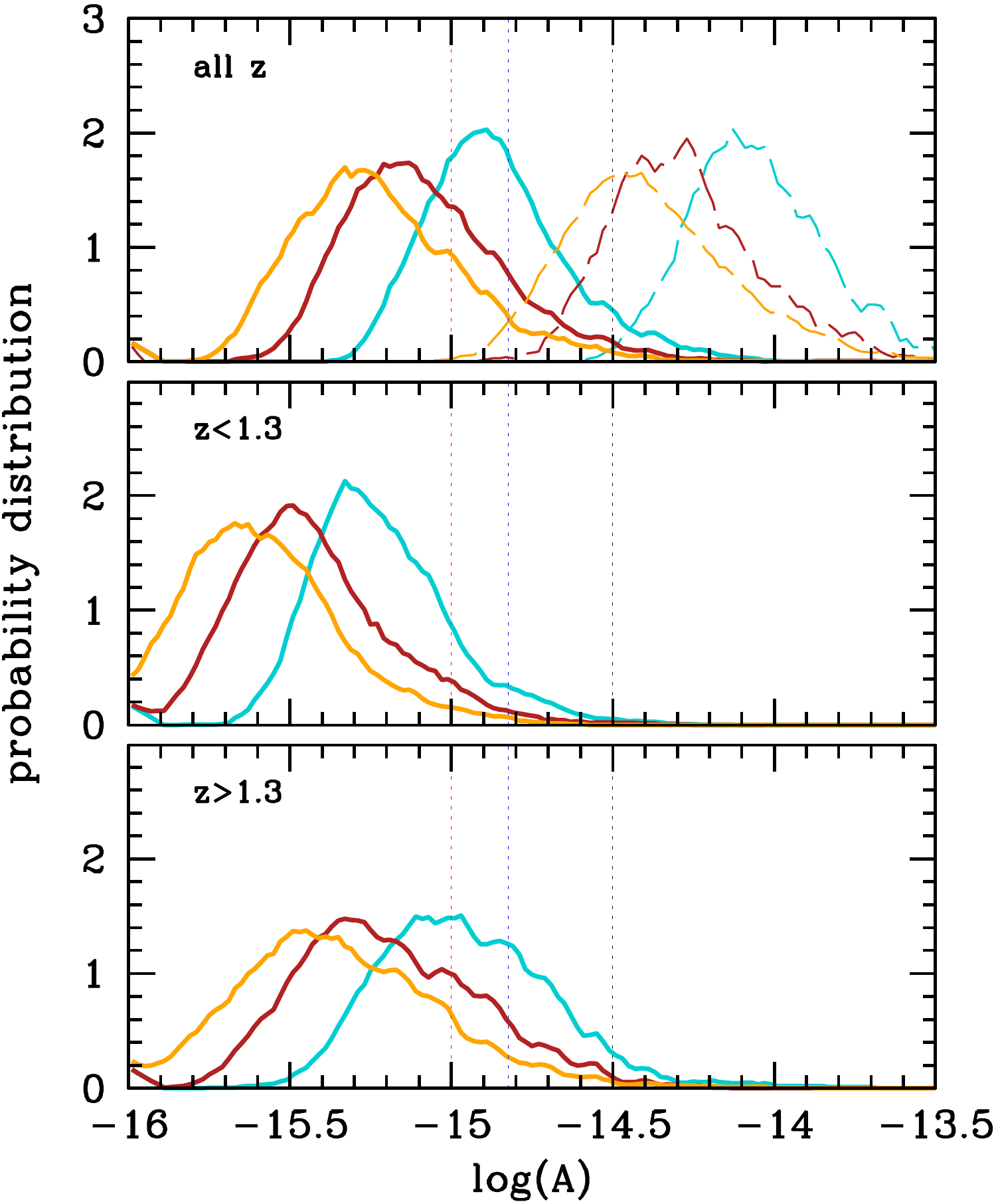}
\caption{Distribution of the stochastic GWB characteristic amplitudes $A$ at the fiducial frequency of 1 yr$^{-1}$ inferred from the CRTS sample under different assumptions about the masses and mass-ratios of the SMBHBs. The blue curves are distributions for Model\_Vir\_hiq (dashed) and Model\_True\_hiq (solid), brown curves are for Model\_Vir\_fid (dashed) and Model\_True\_fid (solid), and orange curves are for Model\_Vir\_loq (dashed) and Model\_True\_loq (solid). The vertical dotted lines are 95\% upper limits given by EPTA (black), NANOGrav (blue) and PPTA (red). The three panels show the overall amplitude distribution considering all CRTS candidates (top) and the contribution of candidates at $z<1.3$ (centre) and $z>1.3$ (bottom). In the latter two panels, only the `True' models are shown.}
\label{fig_a}
\end{figure}
%%%%%%%%%%%%%%%%%%%%%%%%%%%%%%%%%%%%%%%%%
Fig.~\ref{fighc} shows the outcome of the calculation just described for Model\_Vir\_fid. $h_c$ given by equation (\ref{hcdE}) is compared with the expected signal from observation based models featuring different SMBH-galaxy scaling relations. In particular, we pick from \cite{2013MNRAS.433L...1S} a fairly `optimistic' model in which the SMBH mass correlates with the host bulge mass following the relation proposed by \cite{kormendy13}. Note that the GW amplitude predicted by this model, $-15.1<\log A<-14.7$ at 68\% confidence, is already in tension with the current best 95\% GWB upper limit at $\log A<-15$ \citep{2015Sci...349.1522S}. We also consider an alternative model featuring a much more conservative SMBHB-host relation proposed by \cite{2016MNRAS.460.3119S}. This model predicts a GW amplitude of $-15.8< \log A<-15.2$ at 68\% confidence \citep{2016MNRAS.463L...6S}.

The figure highlights the key idea behind our calculation. Although signals from individual candidates are perfectly consistent with current PTA limits, even when virial masses are taken at face value, the implied stochastic GWB extrapolated at lower frequencies is strongly inconsistent with current IPTA upper limits \citep{2015MNRAS.453.2576L,2015Sci...349.1522S,2016ApJ...821...13A}. This is true for all `Vir' models. In the top panel Fig.~\ref{fig_a} we show  the GWB amplitude $A$ at the fiducial frequency of 1 yr$^{-1}$, obtained from the 1000 Monte Carlo realisations of all the 'Vir' models (dashed lines on the right). We obtain $\log(A_{50\%})= -14.11,-14.32,-14.45$ for Model\_Vir\_hiq, Model\_Vir\_fid and Model\_Vir\_loq respectively, which are severely inconsistent with the PTA limits. 

\subsection{Model\_True: lower signal normalization}
%%%%%%%%%%%%%%%%%%%%%%%%%%%%
\begin{table}
\begin{center}
\def\arraystretch{1.5}
\setlength{\tabcolsep}{6pt}
\begin{tabular}{ccccc}
\hline
Name & z & $P$[d] & log$\left(\frac{M_t}{\msun}\right)$ & $A_s[10^{-16}]$\\ %$t_c$[yr]\\
\hline
SDSS J164452.71+430752.2 & 1.715 &  2000 &  10.15 (9.66) & 1.644\\  %110 \\
HS 0926+3608             & 2.150 &  1562 &  9.95 (9.47)  & 0.718\\  %84  \\   
SDSS J140704.43+273556.6 & 2.222 &  1562 &  9.94 (9.45)  & 0.698\\  %84  \\  
SDSS J092911.35+203708.5 & 1.845 &  1785 &  9.92 (9.44)  & 0.698\\  %180 \\
SDSS J131706.19+271416.7 & 2.672 &  1666 &  9.92 (9.45)  & 0.679\\  %77  \\
HS 1630+2355             & 0.821 &  2040 &  9.86 (9.34)  & 0.608\\  %1100\\
SDSS J114857.33+160023.1 & 1.224 &  1851 &  9.90 (9.38)  & 0.544\\  %410 \\ 
SDSS J134855.27-032141.4 & 2.099 &  1428 &  9.89 (9.41)  & 0.515\\  %87  \\
SDSS J160730.33+144904.3 & 1.800 &  1724 &  9.82 (9.37)  & 0.515\\  %250 \\ 
SDSS J081617.73+293639.6 & 0.768 &  1162 &  9.77 (9.28)  & 0.474\\  %360 \\     
\end{tabular}
\caption{Top 10 G15 candidates providing the largest contribution to the expected GWB. Columns give the object identification name, redshift, observed orbital period in days, total mass as given in G15 (median mass estimate when bias is included is given in parenthesis) and individual contribution to the GWB $A_s(f=1{\rm yr}^{-1})$ in Model\_True\_fid.}
\label{tab1}
\end{center}
\end{table}
%%%%%%%%%%%%%%%%%%%%%%%%

Taking virial mass measurements at face value, the CRTS sample is inconsistent with PTA limits, implying that for the majority of the candidates, inferred variability cannot be linked to the presence of a SMBHB. Those mass estimates are, however, known to be biased high, as described in \S~\ref{sec:bias}. We therefore considering the `True' models, where this bias has been corrected via equation (\ref{eq:mtrue}), to achieve more robust conclusions. The GWB amplitude $A$ from 1000 Monte Carlo realizations in each model is shown as solid lines in the top panel of Fig.~\ref{fig_a}, and highlights the critical importance of taking the bias into account. The distributions are now all at least marginally consistent with the most stringent current PTA upper limit \citep{2015Sci...349.1522S} and we obtain $\log(A_{50\%})= -14.88,-15.13,-15.25$ for Model\_True\_hiq, Model\_True\_fid and Model\_True\_loq respectively. The $\approx$0.8dex difference with respect to the `Vir' models is easy to explain. From equation (\ref{hcdE}) $h_c$ is proportional to the square root of the merger rate times the energy spectrum, both of which are proportional to ${\cal M}^{5/3}$. Therefore a reduction of about 0.5dex in the mass estimate (see Fig.~\ref{figbias}) results in a corresponding reduction of $\approx0.8$dex in the GWB. 

Even though the bulk of the GWB is theoretically expected to come from SMBHBs at $z<1.5$ \citep{2008MNRAS.390..192S,2015MNRAS.447.2772R,2016ApJ...826...11S}, the G15 sample features several fairly massive candidates at higher redshifts. Note that, according to equation (\ref{hcdE}), the GW contribution depends on the differential {\it density} of mergers per unit redshift and chirp mass, $d^2n/(dzd{\cal M})$. Each individual candidate in the CRTS sample contributes to this quantity via equations (\ref{eq:tc}) and (\ref{eq:d2ndzdm}). The latter includes the $dz/dV_c$ factor accounting for the comoving volume shell accessible at a given redshift: a single binary observed at higher redshift implies a lower merger rate density because of the larger comoving volume available. One might therefore think that higher $z$ systems do not contribute significantly to the GW signal. Since for a fixed observed frequency binaries at higher redshift are emitting at higher rest-frame frequencies, however, their intrinsic coalescence time is shorter, and the inferred merger rate larger according to equation (\ref{eq:tc}). It turns out that this latter fact compensates for averaging over a larger volume shell, and the GW signal is dominated by SMBHBs at $z>1.3$. This is shown in the centre and bottom panels of Fig.~\ref{fig_a} where we break down the contribution to the GWB from sources at $z<1.3$ and $z>1.3$. The latter contribute, on average, about 2/3 of the total GWB, in striking contrast with theoretical expectations.

\begin{figure}
\includegraphics[scale=0.42,clip=true,angle=0]{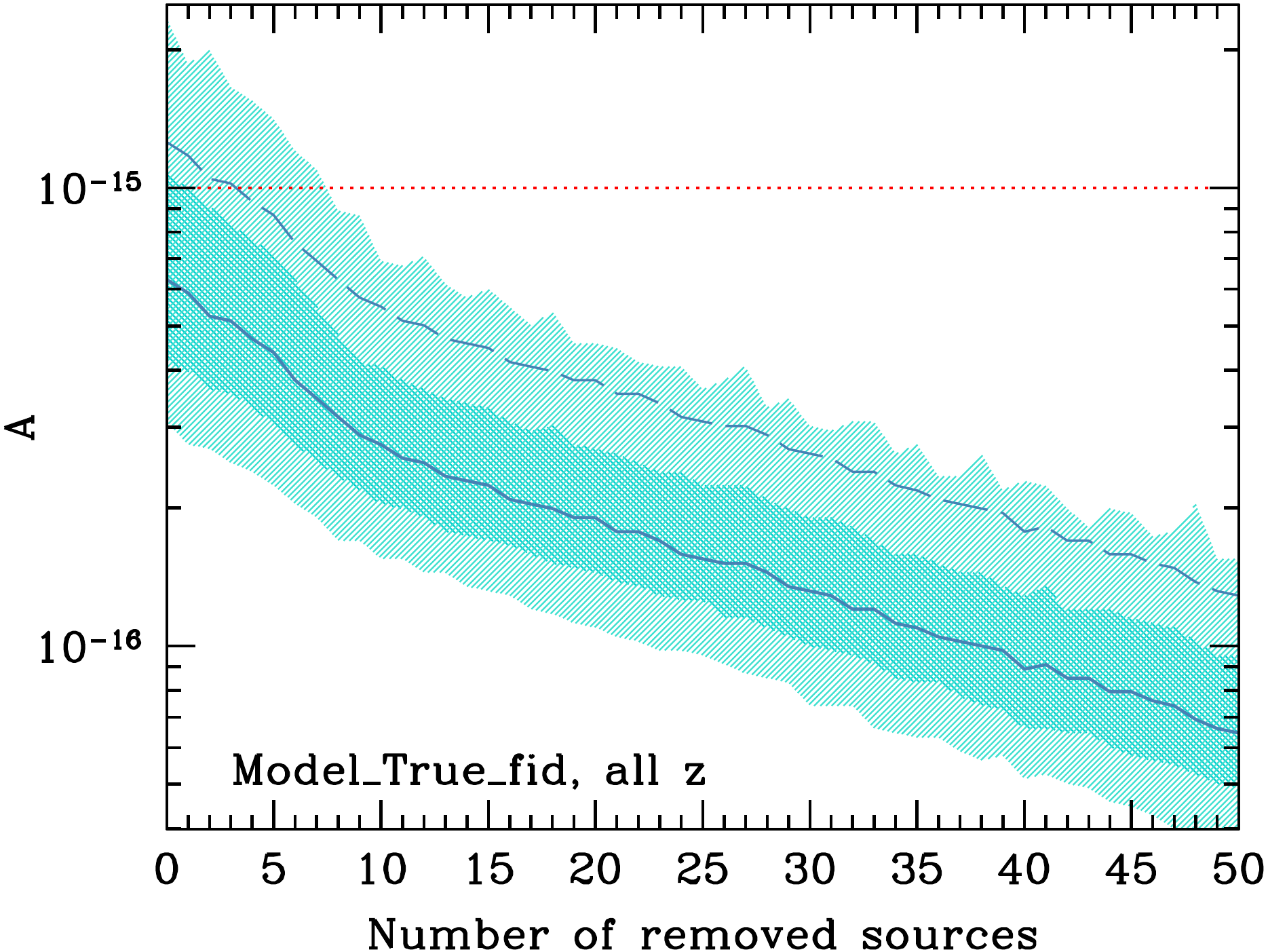}
\caption{Build-up of GWB signal. Plotted is the characteristic strain of the GWB at $f=1$yr$^{-1}$, $A$, versus the number of removed CRTS candidates. Candidates are ranked and removed, starting with the loudest, from left to right. Shaded areas mark the 68\% and 95\% confidence region of the amplitude computed over 1000 realizations of the signal, and the solid blue line is the median value. The dashed blue line is the median shifted upwards by a factor of two to show a conservative estimate of incompleteness (the real impact of incompleteness is likely larger; see discussion in \S~\ref{sec:incompleteness}). The horizontal dotted red line is the PPTA upper limit. Model\_True\_fid is assumed.}
\label{fig_buildup}
\end{figure}

To single out the individual candidates contributing the most to the GWB, we performed the following experiment. Under the assumption of the fiducial Model\_True\_fid, we generated 1000 realisations of $M_1$ and $M_2$ of each of the 98 CRTS binaries. We then used equations (\ref{eq:tc}) and (\ref{eq:d2ndzdm}) to compute their contribution to the SMBHB merger density, and folded the result into equation (\ref{hcdE}) to compute the associated GW signal. We stress again that although the merger rate obtained via equation (\ref{eq:dndtr}) depends on the chosen bin size, this is compensated by the integral over the size of the bin in equation (\ref{hcdE}). We then rank the sources by individual contribution to the GWB for each realization of the signal, progressively removing sources one by one from the loudest to the quietest. The result is shown in Fig.~\ref{fig_buildup} for the progressive removal of the loudest 50 systems. The overall signal drops steeply as the first 10 sources are removed. Their median contribution to the GWB for this specific model is listed in the last column of  Tab.~\ref{tab1}. Note that their combined contribution alone accounts for about 70\% of the overall GWB and note also that the five loudest candidates are at $z>1.5$, where we do not expect a significant contribution to the GWB from theoretical models. This analysis puts the CRTS candidate distribution strongly at odds with theoretical expectations, even while the distribution is marginally consistent with PTA observations alone.

%%%%%%%%%%%%%%%%%%%%%%%%%%%%%%%%%%%%%%%%%
%%%%%%%%%%%%%%%%%%%%%%%%%%%%%%%%%%%%%%%%%

\subsection{Correcting for incompleteness: tension between Model\_True and PTA upper limits}
\label{sec:incompleteness}
%%%%%%%%%%%%%%%%%%%%%%%%%%%%%%%%%%%%%%%%%
\begin{figure}
\includegraphics[scale=0.45,clip=true,angle=0]{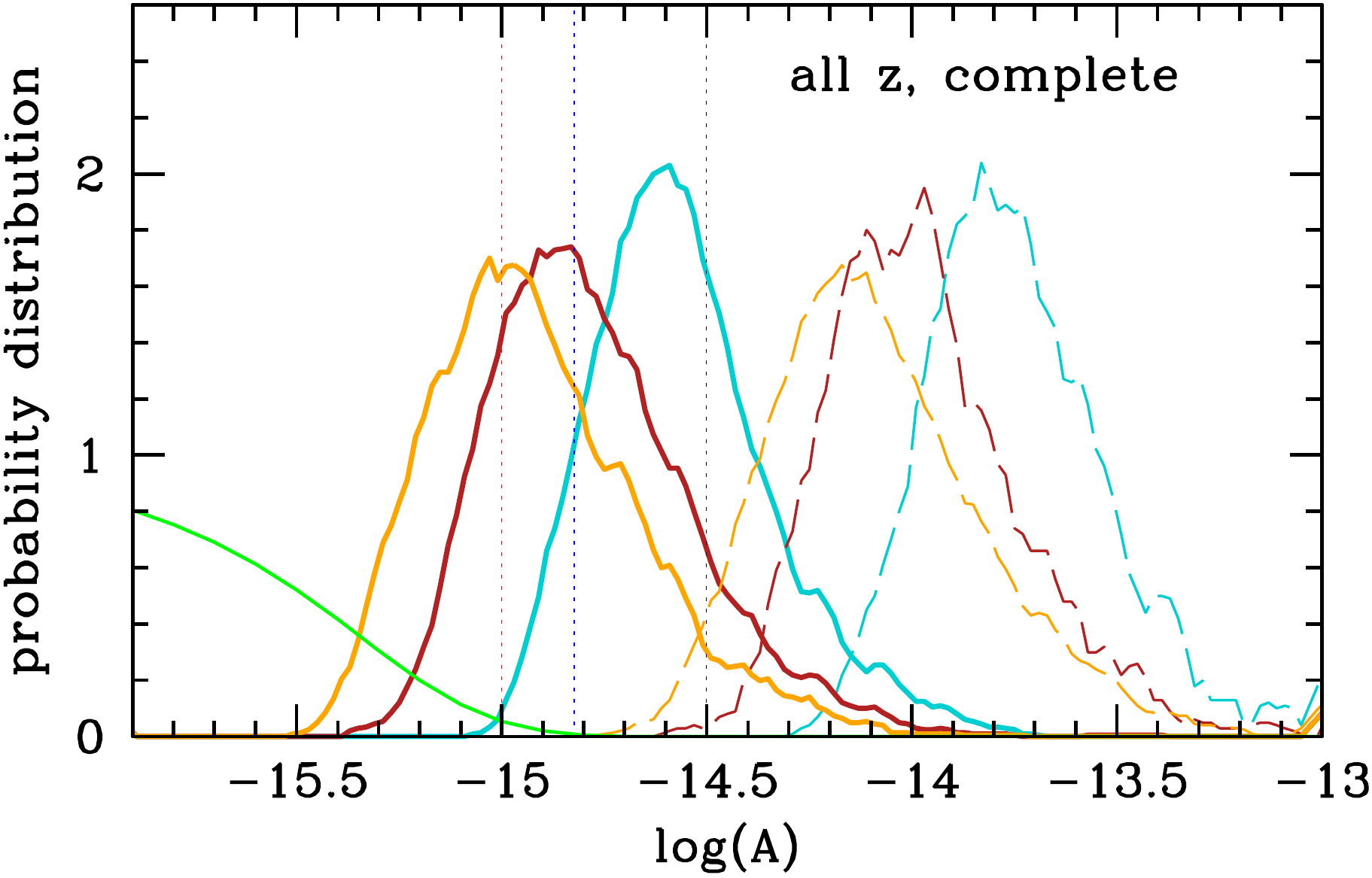}
\caption{Same as the top panel of Fig.~\ref{fig_a}, but correcting for completeness of the CRTS sample. Line style as in Fig.~\ref{fig_a}, the extra green line represents the posterior sample distribution of the PPTA amplitude measurement from which the 95\% upper limit of $A=10^{-15}$ is derived.}
\label{money1}
\end{figure}
%%%%%%%%%%%%%%%%%%%%%%%%%%%%%%%%%%%%%%%%%

Although the CRTS sample seems consistent with current PTA upper limits, we did not yet consider the fact that the SMBHB sample is necessarily incomplete in several ways. First, CRTS covers only 80\% of the sky (about 33000deg$^2$). Although this is a significant fraction of the whole sky, most of the selected quasars are identified in the SDSS survey (78 out of 111), which covers only about one quarter of the sky. Assuming a complete identification in the SDSS field of view, we would therefore expect $78\times4=312$ objects in the whole sky. We conclude that the G15 catalogue is at best only 35.5\% complete based on sky coverage arguments only. Moreover, of the 334k identified quasars, 83k (25\%) were rejected because of sparse sampling. Finally, we dropped 13 candidates with no mass estimate (12\%) from the sample. Adding everything up, the sample we use is {\it at best} only 24\% complete. Since the signal is proportional to the square root of the coalescence rate, all the signal estimates shown so far must be multiplied by {\it at least} a factor of two.

The effect of incompleteness is shown in Fig.~\ref{money1}. The GWB predicted from the G15 sample, corrected for incompleteness, starts now to be in tension with current PTA measurements, in particular with the PPTA limit. In fact, we find $\log(A_{50\%})= -14.60,-14.82,-14.97$ for Model\_True\_hiq, Model\_True\_fid and Model\_True\_loq respectively, all of which are above the 95\% upper limit at $\log(A)= -15$, published by PPTA. When virial mass measurements are assumed (`Vir' models), the predicted amplitude is severely inconsistent with all PTA upper limits.

Note that, besides these simple `counting arguments' which can be easily corrected for, there are potentially many more sources of incompleteness in the G15 SMBHB sample. First, the redshift distribution of the CRTS candidate quasars shows a prominent peak around $z\approx0.8$, hinting to incompleteness at higher redshifts. Second, the candidates are all spectroscopically confirmed type 1 quasars; numbers should therefore be corrected for the fraction of obscured type 2 quasars, which is poorly known but can easily double the sample \citep{2013ApJ...777...86L}. Third, the search is constructed to look for sinusoidal periodicities and would be much less sensitive to eccentric binaries. Finally, not all SMBHBs have to be active in the first place, especially in gas poor mergers that are rather frequent at low redshift. In any case, the factor of two-corrected signal shown in Fig.~\ref{money1} is necessarily a lower limit to the intrinsic GWB implied by the CRTS SMBHB candidate population.

\subsection{Bayesian model selection: preference for the null hypothesis}
\label{sec:modelsel}
%%%%%%%%%%%%%%%%%%%%%%%%%%%%%%%%%%%%%%%%%%%%%%%%%%%%%%%
\begin{table}
\begin{center}
\begin{tabular}{c|ccc}
\hline
\multicolumn{1}{c|}{model pair} & \multicolumn{3}{c}{odds ratio \& probabilities}\\
              & log$\Lambda_{NB}$ & $p_N$ & $p_B$\\
\hline
{\rm N/True\_hiq} & $2.61$ & $0.9976$  & $0.0024$\\ 
{\rm N/True\_fid} & $1.52$ & $0.9709$  & $0.0291$\\ 
{\rm N/True\_loq} & $1.10$ & $0.9267$  & $0.0733$\\ 
{\rm N/Vir\_hiq}  & $11.1$ & $>0.9999$ & $<10^{-11}$\\ 
{\rm N/Vir\_fid}  & $6.67$ & $>0.9999$ & $<10^{-6}$\\ 
{\rm N/Vir\_loq}  & $5.22$ & $>0.9999$ & $<10^{-5}$\\ 
\hline
\end{tabular}
\end{center}
\caption{Model selection results: pair comparisons between the null hypothesis and all the investigated models based on the CRTS sample. For each two--model comparison, we report the log of the likelihood ratio $\Lambda_{NB}$, and the probability of the null hypothesis ($p_N$) and the binary hypothesis ($p_B$).}
\label{tab2}
\end{table}
%%%%%%%%%%%%%%%%%%%%%%%%%%%%%%%%%%%%%%%%%%%%%
We now wish to properly quantify the concept of `tension' between the CRTS sample and the PTA measurements. To do so, we employ the concept of Bayesian model selection considering two competitive hypotheses: the null hypothesis {\it N} that the CRTS candidates are not binaries, versus the hypothesis {\it B} that the candidates are indeed binaries. When considering non-parametric models, the odds ratio of model {\it N} over model {\it B} is simply given by,
\begin{equation}
  \Lambda_{NB}=\frac{p(D|N)}{p(D|B)}\frac{p(N)}{p(B)},
\end{equation}
where the model likelihood $p(D|X)$ is the likelihood of the observed data $D$ under the $X$ hypothesis, and $p(X)$ is the prior probability assigned to model $X$. If we make the agnostic assumption that models {\it N} and {\it B} are a priori equally probable, then the odds ratio reduces to computing the likelihood ratio of model {\it N} over {\it B}. In a pairwise comparison it is then possible to associate a probability $p_N=p(D|N)/(p(D|N)+p(D|B))$ to the null hypothesis, and a probability $p_B=1-P_N$ to the binary hypothesis.

Given the data, the model likelihoods can be evaluated as the integral of the amplitude distribution predicted by a specific model times the posterior amplitude distribution derived by the data. The likelihood of model {\it X} is thus:
\begin{equation}
  p(D|X)=\int {\cal F}(A)p_X(A){\rm d}A.
  \label{eq:like}
\end{equation}
We consider here the PPTA data used in \cite{2015Sci...349.1522S}. The posterior amplitude distribution is fairly well described by a Fermi function of the form,
\begin{equation}
  {\cal F}(A)=\frac{C_1}{e^{(A-C_2)/C_3}+1},
  \label{eq:fermi}
\end{equation}
with $C_1=1.63$, $C_2=1.2\times10^{-16}$ and $C_3=2.6\times10^{-16}$ \citep{2017arXiv170700623M}, and is shown in Fig.~\ref{money1}. We need now to specify the amplitude probability of the competing models. In the null hypothesis, no binaries are present in the data, the amplitude distribution is therefore a delta function centred at $A=0$, $p(A)=\delta_0$. We consider pairwise comparisons with each of the binary models proposed in this paper, so model {\it B} will be  Model\_True\_hiq, Model\_True\_fid, Model\_True\_loq, Model\_Vir\_hiq, Model\_Vir\_fid, Model\_Vir\_loq and the associated amplitudes are those computed from the 1000 realizations of the GWB, corrected for incompleteness in each case, as shown in Fig.~\ref{money1}. 

%%%%%%%%%%%%%%%%%%%%%%%%%%%%%%%%%%%%%%%%%
\begin{figure*}
\begin{tabular}{cc}
\includegraphics[scale=0.42,clip=true]{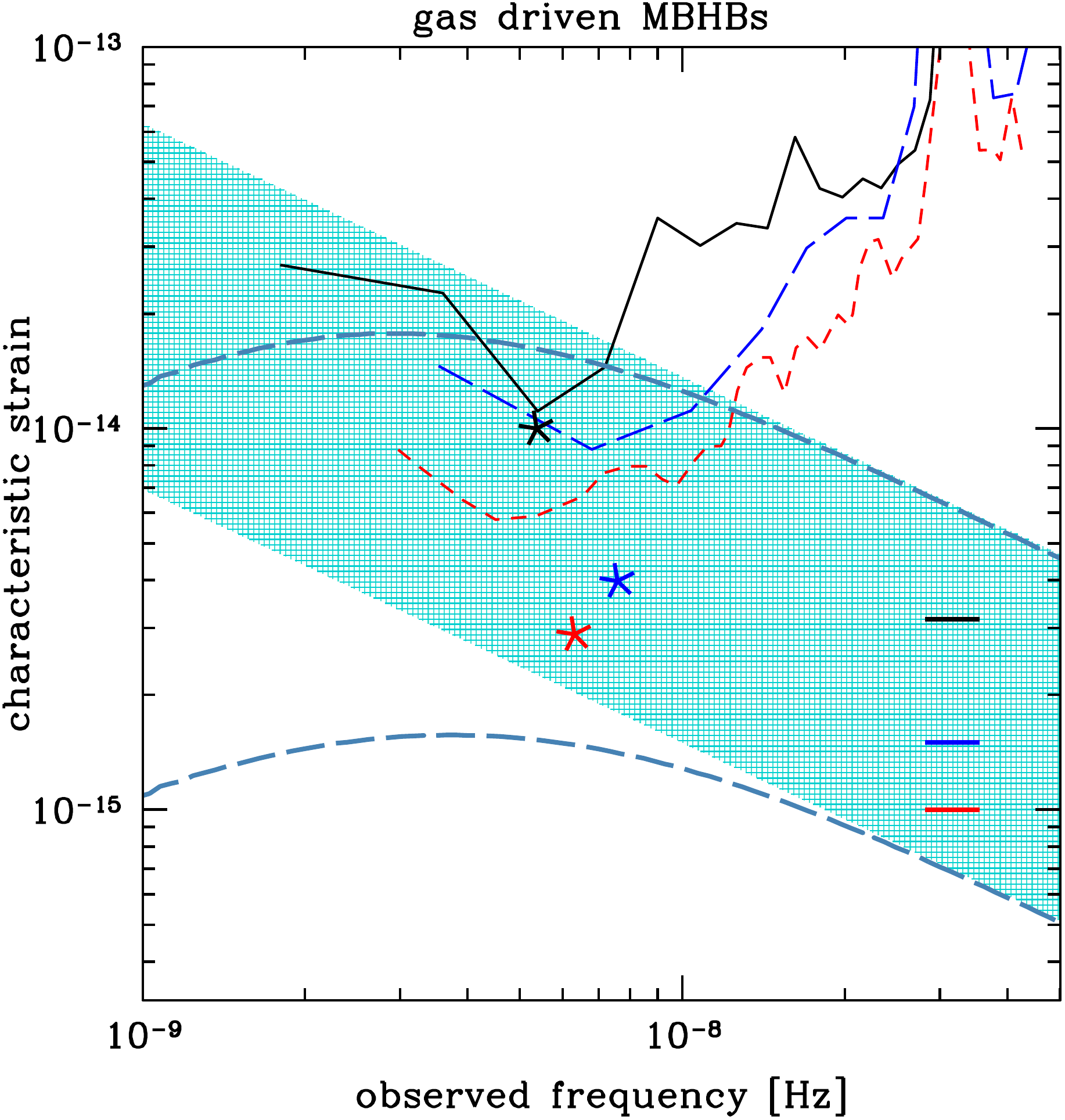}&
\includegraphics[scale=0.42,clip=true]{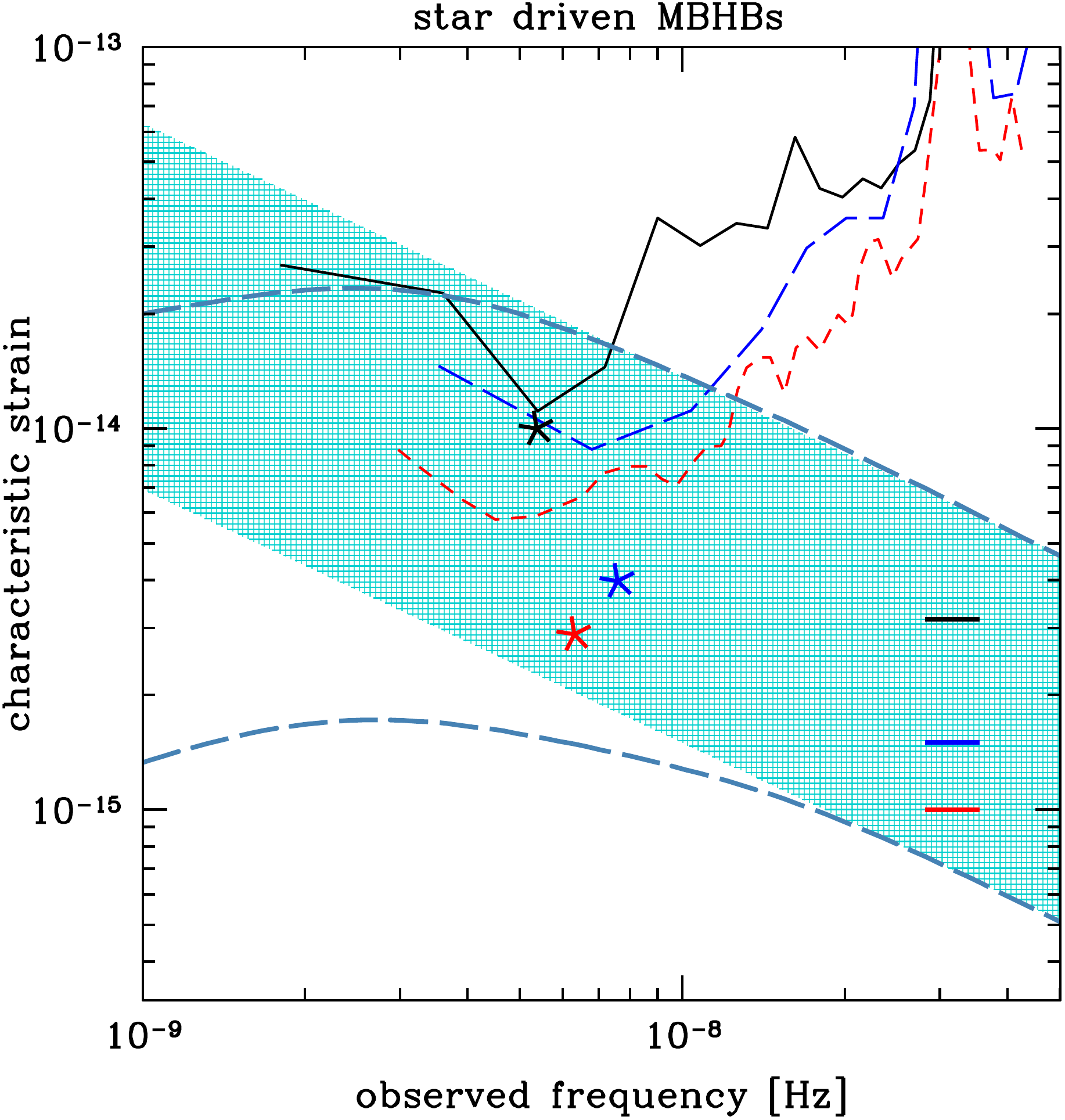}\\
\end{tabular}
\caption{Effect of gas (left) and stellar (right) dynamics on the GW signal implied by the CRTS SMBHB sample (see text for details of the models of SMBHB-environment couplings). In each panel, the blue shaded area is the 95\% confidence intervals produced by the fiducial Model\_True\_fid. The area was computed including G15 SMBHBs at all redshifts and correcting for incompleteness of the sample (see text for full details). The long-dashed blue lines represent the 95\% predicted range for $h_c$ under the effect of environmental coupling. The representation of the PTA sensitivity curves and upper limits is the same as in Fig.~\ref{fighc}.}
\label{fighcgs}
\end{figure*}
%%%%%%%%%%%%%%%%%%%%%%%%%%%%%%%%%%%%%%%%%

Note that ${\cal F}(0)=1$ so that the likelihood of the null hypothesis is trivially $p(D|N)=1$, whereas the likelihood of the {\it B} hypothesis has to be computed via numerical integration of equation (\ref{eq:like}). Results are shown in Tab.~\ref{tab2}. It is clear that the `Vir' models are inconsistent with PTA limits and are in fact strongly disfavoured compared to the null hypothesis. The situation is not so conclusive when the fiducial `True' models are considered. log$\Lambda_{NB}$ are now in the range 1.1--2.5, indicating substantial preference for the null hypothesis \citep{kassr95}. Depending on the mass ratio distribution, the null hypothesis is preferred at the 99.7\%, 97.1\% and 92.7\% confidence level. Translated in the familiar `$\sigma$-level' jargon, the null hypothesis is preferred over our fiducial model (Model\_True\_fid) at $\approx2.3\sigma$. 

\subsection{Possible impact of SMBHB coupling with stars and gas}
\label{coupling}

In the previous discussion, we considered circular GW-driven binaries. However, it has been shown that both coupling with the environment and large eccentricities can in principle suppress the GW signal at nHz frequencies \citep[see e.g.][]{2007PThPh.117..241E,2011MNRAS.411.1467K,2013CQGra..30v4014S,2014MNRAS.442...56R}.

Since the G15 SMBHB candidates are quasars, it is important to explore the case of strong coupling with a gaseous environment. Relevant models were constructed by \cite{2011MNRAS.411.1467K}. They studied the SMBHB-disc coupling assuming different thin disc models \citep{1973A&A....24..337S} described in \cite{2009ApJ...700.1952H}. In particular they considered steady-state solutions by \cite{1995MNRAS.277..758S}, assuming both $\alpha$ and $\beta$ discs, and the self-consistent non-steady solution of \cite{1999MNRAS.307...79I}. They found that only $\alpha$ discs with a large viscosity parameter ($\alpha_{SS}=0.3$, in that work) and large accretion rate ($\dot{M}/\dot{M}_{\rm Edd}=0.3$, in that work) can significantly suppress the signal at the frequencies where current PTAs are sensitive. We therefore consider this model here, even though it might be unlikely to occur in nature because the secular thermal-stability of $\alpha$-discs is uncertain \citep[see, e.g.,][]{2009ApJ...704..781H,2013ApJ...778...65J}. Moreover, we include eccentricity, according to the findings of \cite{2011MNRAS.415.3033R}, where binaries have $e\approx 0.6$ as long as they are coupled with the disc, and they progressively circularize due to GW emission after decoupling. Results are shown in the left panel of Fig.~\ref{fighcgs} for Model\_True\_fid. Even considering such an extreme coupling, the implied GW signal is reduced only by a factor of $\approx 1.5$ at 6 nHz (and $\approx 7$ at 1 nHz) and is still generally in tension with PTA upper limits. We stress that any other model (i.e.~alpha discs with a smaller $\alpha$-viscosity, any $\beta$ disc, or any non-steady self-consistent disc) would result in a negligible suppression of the signal in the PTA frequency range. 

Coupling with stars can also accelerate SMBHB evolution and promote eccentricity growth \citep[see, e.g.,][]{1996NewA....1...35Q,2010ApJ...719..851S}, both of which suppress the low frequency GW signal. \cite{2010ApJ...719..851S} modelled the stellar environment as a broken power-law, with a central cusp of density $\rho\propto r^{-\gamma}$ joining an external isothermal sphere at the SMBHB influence radius. In those models the SMBHB loss cone is assumed to be always full, which implies that the binary evolves at a pace that is dictated by the stellar density $\rho_0$ at the influence radius. Since the isothermal sphere is quite compact, especially compared to observed inner density profiles of massive ellipticals, these models can overestimate $\rho_0$ by more than an order of magnitude, and are therefore aggressive in terms of the efficiency of the SMBHB-stellar coupling. We consider here the model with $\rho\propto r^{-1}$ and $e_0=0.6$ ($e_0$ is the eccentricity at binary formation). Results are shown in the right panel of Fig.~\ref{fighcgs} again for Model\_True\_fid. Also in this case, the suppression of the GWB is mild, and the result is still in tension with PTA upper limits.

In principle, a larger suppression is possible if SMBHBs are more eccentric. However, as an example, the Bayesian analysis performed by \cite{2015Natur.525..351D} on one of the G15 binary candidates, the $z=0.3$ quasar PG~1302-102, showed that its eccentricity is at most $\approx 0.2$ and it is consistent with zero. All SMBHBs in CRTS have been identified by their sinusoidal behaviour, which implies small eccentricities. The models considered in this section are consistent with SMBHBs having $e<0.25$ at $f>10^{-8}$Hz, i.e.~in the frequency range probed by CRTS. For example, a stellar driven model with $e_0=0.8$ would result in a population of fairly eccentric binaries at $f>10^{-8}$Hz \citep[e.g.][]{2017MNRAS.471.4508K}, inconsistent with the sinusoidal light curves observed in the CRTS sample.

We caution that our computation {\it underestimates} the signal in the case of extreme environmental couplings. In fact, the normalization of the signal is set by the merger rate that is computed according to equation (\ref{eq:dndtr}) assuming circular GW-driven binaries. If SMBHBs are still coupled with the environment, then the coalescence time $T_c(f)$ would be shorter, resulting in a higher merger rate and, in turn, in a higher signal normalization. 

\section{Application to the PTF sample}
\label{sec5}
We next apply the same methodology described in \S~\ref{sec3} to the PTF sample identified by C16. By studying a sample of 35,383 spectroscopically confirmed quasars from PTF, they construct a sample of 33 binary candidates with periods $\lesssim 500$ days. Masses reported in C16 are either from \cite{2008ApJ...680..169S} or estimated by the authors using a similar technique based on the luminosity alone (in those cases when  \cite{2008ApJ...680..169S} measurements were unavailable; note that the luminosity-based estimates have a slightly larger uncertainty). These are therefore again virial masses, and a systematic bias, as in the CRTS sample, should be present. We therefore consider also in this case the same `Vir' and `True' models, constructed as described in \S~\ref{sec:bias}, assuming $\tilde{P}_{\rm orb}=500$days, consistent with the shorter time-span of the PTF dataset.

We also note that the sample is affected by severe incompleteness. Of the 33 candidates, 28 fall in the SDSS footprint. Considering that the PTF and SDSS footprints overlap over 9700deg$^2$, one would expect $28\times (41253/9700)=119$ systems for a uniform sky coverage. Moreover, of the 278,740 spectroscopically confirmed quasars in PTF, only 35,383 passed the selection criteria in terms of observational cadence. Considered together, these two facts imply that the C16 sample is only $\lesssim$3\% complete. This implies an upward correction factor of $\approx \sqrt{40}\approx 5.8$ when computing the associated GWB.

\begin{figure}
\includegraphics[scale=0.42,clip=true,angle=0]{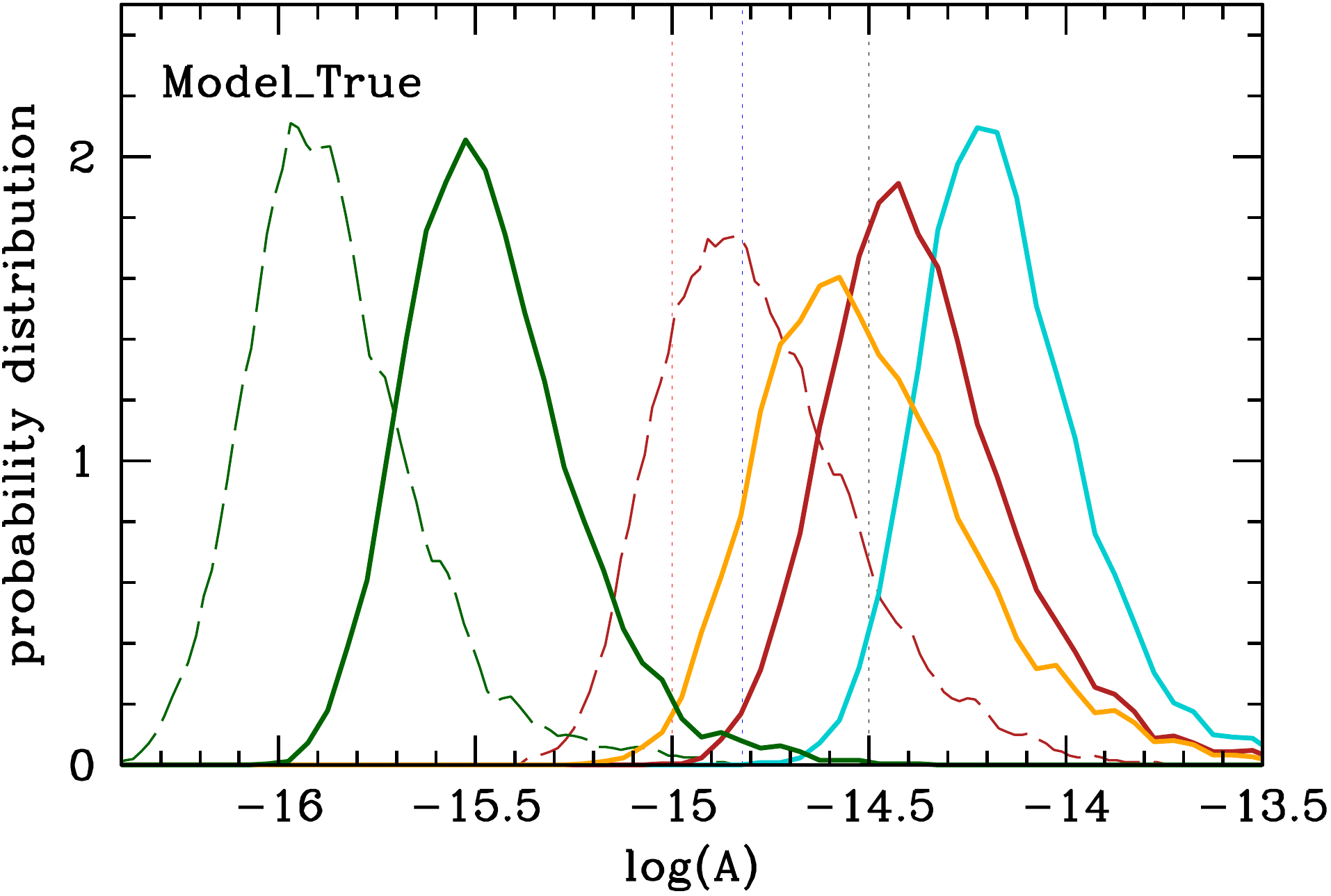}
\caption{Distribution of the stochastic GWB characteristic amplitudes inferred from the PTF sample. Shown are the default Model\_True\_hiq (light-blue), Model\_True\_fid (brown) and Model\_True\_loq (orange) alongside with an extreme model in which all candidates have $q=0.01$ (thick solid green). Thin dashed lines show for comparison the GWB produced by the CRTS sample under Model\_True\_fid (brown) and assuming $q=0.01$ (green). The vertical dotted lines are 95\% upper limits given by EPTA (black), NANOGrav (blue) and PPTA (red).}
\label{fig_charisi}
\end{figure}

We find that, when corrected for incompleteness, the predicted GWB amplitudes in the `Vir' models are inconsistent with PTA upper limits by more than an order of magnitude. `True' models are also inconsistent with the data, as shown in Fig.~\ref{fig_charisi}. The average GWB amplitude is estimated to be $\log(A_{50\%})= -14.20,-14.42,-14.57$ for Model\_True\_hiq, Model\_True\_fid and Model\_True\_loq respectively. If we apply the analysis described in \S~\ref{sec:modelsel} those models are inconsistent with the PPTA measurement at a level of $4.5\sigma$, $3.6\sigma$ and $3.0\sigma$ as reported in Tab.~\ref{tab3}.

When analyzing this sample, Charisi and co-authors find that the observed period distribution best matches the expected SMBHB coalescence time distribution if their mass ratio is $q\approx 0.01$. We therefore constructed an alternative model where we assumed `True' mass estimates but an extreme mass ratio $q=0.01$ for all candidates. The expected GWB amplitude (after correcting for incompleteness) is also displayed in Fig.~\ref{fig_charisi} in green. Under this ansatz, the PTF sample would be consistent with current PTA data. Note that the same would be true for the CRTS sample (also shown in the figure).

A few things should be noted though. A mass ratio distribution of $q\approx 0.01$ might be expected if the periodicity is due to Doppler boosting, as also noted by C16. If this is the case, however, there is an observational bias in favour of SMBHBs with small inclination angle (i.e. nearly edge on). A proper estimate of the GWB within this interpretation should take this bias into account.  Moreover, we did not consider any correction for missing SMBHBs with $q>0.01$; if Doppler boosting manifests itself only at these small $q$, then there must be a large population of undetected SMBHBs with larger $q$ contributing to the GWB, likely boosting it to unacceptable levels. Alternatively, one has to envisage a model whereby only $q\approx 0.01$ binaries form, which would be unexpected in the context of hierarchical build-up of SMBHs.    

Summarizing:
\begin{itemize}
\item the PTF sample is at 3\% complete;
\item after correcting for incompleteness, all Model\_Vir are more than an order of magnitude inconsistent with PTA upper limit;
\item the fiducial model Model\_Vir\_fid is inconsistent with the PPTA upper limit at $3.6\sigma$;  
\item if we assume $q\approx 0.01$ as in C16, then the PTF sample is consistent with PTA data. However the GWB computed in this way is necessarily a lower limit to the total GWB because of: i) inclination bias and ii) higher $q$ binaries being missed in the sample. 
\end{itemize}

%%%%%%%%%%%%%%%%%%%%%%%%%%%%%%%%%%%%%%%%%
%%%%%%%%%%%%%%%%%%%%%%%%%%%%%%%%%%%%%%%%%
%%%%%%%%%%%%%%%%%%%%%%%%%%%%%%%%%%%%%%%%%%%%%%%%%%%%%%%
\begin{table}
\begin{center}
\begin{tabular}{c|ccc}
\hline
\multicolumn{1}{c|}{model pair} & \multicolumn{3}{c}{odds ratio \& probabilities}\\
              & log$\Lambda_{NB}$ & $p_N$ & $p_B$\\
\hline
{\rm N/True\_hiq} & $5.55$ & $>0.9999$  & $<10^{-5}$\\ 
{\rm N/True\_fid} & $3.60$ & $0.9997$  & $0.0003$\\ 
{\rm N/True\_loq} & $2.57$ & $0.9973$  & $0.0027$\\ 
\hline
\end{tabular}
\end{center}
\caption{Same as Tab.~\ref{tab2} for the PTF sample. Only `True' models are displayed.}
\label{tab3}
\end{table}
%%%%%%%%%%%%%%%%%%%%%%%%%%%%%%%%%%%%%%%%%%%%%

\section{Conclusions}
\label{sec6}

We have shown that the list of supermassive black hole binary candidates identified by \cite{2015MNRAS.453.1562G} in the CRTS and by \cite{2016MNRAS.463.2145C} in the PTF are both in tension with current pulsar timing array limits on a stochastic gravitational wave background at nHz frequencies. The bulk of our analysis focused on the CRTS sample. Although none of the candidates, taken individually, is inconsistent with PTA measurements, they can be collectively used to construct a cosmic SMBHB merger rate and the associated GWB at nHz frequencies. The GWB computation implies the knowledge of the chirp masses of the candidates; we therefore need to assign a total mass and a mass ratio to each object. We considered two mass models that we labelled Model\_Vir and Model\_True. In the former, virial masses are directly taken from  \cite{2008ApJ...680..169S}; whereas in the latter, prior knowledge of the SMBH mass function is used to infer the `true' mass of each candidate from its virial mass, thus correcting the intrinsic bias in individual virial mass estimates. For each of the mass models we explored three different possible mass ratio (q) distributions: one preferentially high (hiq), another low (loq), and a fiducial intermediate model (fid). This gives a total of six models: Model\_True\_hiq, Model\_True\_fid, Model\_True\_loq, Model\_Vir\_hiq, Model\_Vir\_fid, Model\_Vir\_loq.    

Our main findings can be summarized as follows:
\begin{enumerate}
\item The GWB calculation critically depends on the mass estimates of the SMBHB candidates. Each Model\_Vir results in a GWB that is $\approx$0.8dex larger then the corresponding Model\_True.
\item The mass ratio distribution has a milder impact, affecting the GWB at a factor of $\approx 2$ level. This is because the GW signal depends on the chirp mass, ${\cal M}=Mq^{3/5}$, that has a milder dependence on $q$.  
\item All Model\_Vir are severely inconsistent with all PTA measurements. This is, however, understandable since virial mass estimates are known to be biased high. 
\item When corrected for sample incompleteness, all Model\_True show significant tension with the PPTA upper limit (the most stringent to date), at a level that ranges between $1.8\sigma$ and $3\sigma$ depending on the mass ratio distribution. Our fiducial Model\_True\_fid is inconsistent with the PPTA measurement at $2.3\sigma$. 
\item About $2/3$ of the GWB is generated by sources at $z>1.3$ in stark contrast with theoretical expectations; this is likely due to the presence of selection effects in the CRTS sample.
\item On the other hand, about 70\% of the signal is contributed by a set of 10, mostly high-redshift SMBHB candidates (\cf Fig.~\ref{fig_buildup} and Tab.~\ref{tab1}). Misidentification of a few particularly loud systems as binaries can therefore have a significant impact on the expected GWB.
\item Environmental coupling can decrease the signal by at most a factor of 1.5 at 6 nHz (i.e.~where current PTAs are most sensitive), only partly alleviating the tension with PTA measurements. We stress that this estimate is optimistic, because it does not take into account that coupling with the environment which would shorten the coalescence time scale of the SMBHB candidates, resulting in a higher merger rate and, in turn, in a higher GWB normalization (which we ignored in \S~\ref{coupling}).
\end{enumerate}

We stress that we implicitly assumed that {\it all} SMBHBs with orbital periods of few years are periodic quasars. If only a fraction of those SMBHBs exhibit AGN activity than the predicted stochastic GW background would severely violate the current PTA upper limits.

These results show that the SMBHB hypothesis for the full CRTS sample, as presented in G15, is in tension with PTA measurements. To get some insight on how this candidate sample can be reconciled with current PTA upper limits, we turn to equation (\ref{hcdE}). This equation shows that $h_c$ is proportional to the square root of the SMBHB merger rate, which is in turn proportional to the number of SMBHB candidates as per equation (\ref{dndt}). An average factor of $\approx$2 suppression (required to make our fiducial Model\_True\_fid fully consistent with PTA data) is therefore achievable if at least  75\% of the systems are not SMBHBs. Note, however, that the contribution to the GWB varies significantly across candidates (see Fig.~\ref{fig_buildup} and Tab.~\ref{tab1}). If the five loudest systems are excluded, the GWB implied by the rest if the sample would be consistent with current PTA upper limits even after incompleteness correction (cf Fig.~\ref{fig_buildup}). 

If we want to leave the number of SMBHB candidates untouched, the other parameter to look into is the inferred mass of each system.  Equation (\ref{hcdE}) implies in this case that $h_c\propto{\cal M}^{5/3}$. A factor ${\cal M}^{5/6}$ comes from the square root of $dE/d$ln$f$ and another ${\cal M}^{5/6}$ contribution comes from equations (\ref{eq:tc}) and (\ref{dndt}). If the candidate masses are lower, the coalescence time scale is longer and the implied merger rate smaller. So in principle, a mild reduction of the true mass estimate by a mere factor of 1.5 would suffice to reconcile the CRTS sample with PTA data. Alternatively, the mass ratio distribution of all these systems can be severely biased towards $q\ll 1$. However, a scenario in which all SMBHBs have $q\ll 1$ is difficult to accommodate in current galaxy formation models. Moreover, it is not clear whether low mass-ratio perturbers would result in a luminosity modulation as large as 50\%, as observed in the CRTS sample \citep{Dorazio+2016}.

Alternatively, it has been proposed that the optical variability of SMBHB may be related to the appearance of an $m=1$ mode at the inner edge of the circumbinary disk \citep{2015MNRAS.452.2540D,2017MNRAS.466.1170M}. In this interpretation, however, the binary orbital frequency would be $\approx 5$ times higher than the observed variability in the CRTS sample, implying typical coalescence rates a factor $5^{8/3}\approx 70$ larger and an associated GWB louder by almost an order of magnitude. It is clear that this interpretation of the CRTS candidates is not viable. Conversely, C16 mentioned the possibility that the periodicity might be related to higher harmonics of the orbital periods. This would of course imply much longer periods and coalescence times, making PTA GWB upper limits less constraining.

A similar analysis applied to the PTF sample identified by C16, yields comparable results. When the sample is corrected for incompleteness and virial mass bias, the resulting GWB is inconsistent with PTA upper limits at  $4.5\sigma$, $3.6\sigma$ and $3.0\sigma$, depending on the postulated $q$ distribution. Conversely, the alternative model proposed by C16, whereby all candidates have $q\approx 0.01$, is consistent with current PTA constraints. This is justified under the assumption that periodicity is due to Doppler boosting, which introduces a bias towards such small $q$. We note, however, that in this interpretation the GWB has to be corrected for the missing fraction of SMBHBs with larger mass ratios. Alternatively, one has to put forward a model whereby only binaries with $q\approx 0.01$ form, which would be unexpected in the current framework of SMBH assembly.

In summary, we conclude that both the CRTS and the PTF candidate samples are in moderate tension with the current PTA measurements. Possibilities to alleviate this tension preserving the binary hypothesis include: (i) even after being corrected for the known virial mass estimate bias, typical true SMBH masses have been overestimated by a factor of ($\gsim 1.5$), (ii) the typical mass ratio is lower than theoretically expected ($q \lsim 0.1$), or else (iii) that the loudest GW sources are preferentially false positives.

While these results question the viability of SMBHB identification via periodicity studies alone, they demonstrate the status of PTAs as important astrophysical probes. Even without a direct GWB detection, PTA upper limits can put stringent constraints on interesting candidate objects. Further studies of these systems are required to identify the true origin of their periodic variability. 

\acknowledgments{We thank Scott Tremaine and Maria Charisi for useful discussions. AS is supported by a University Research Fellowship of the Royal Society.  Financial support was provided by NASA through ATP grant NNX15AB19G and ADAP grant NNX17AL82G and by NSF grant 1715661 (ZH). ZH also gratefully acknowledges support from a Simons Fellowship in Theoretical Physics. This project has received funding from the European Research Council (ERC) under the European Union's Horizon 2020 research and innovation programme (grant agreement No 638435) (GalNUC) and from the Hungarian National Research, Development, and Innovation Office grant NKFIH KH-125675 (to BK).}

\bibliographystyle{yahapj}
\bibliography{references}

\end{document}